\documentclass[11pt]{article}
\newcommand{\pa}{\partial}
\newcommand{\be}{\begin{equation}}
\newcommand{\ee}{\end{equation}}
\newcommand{\bea}{\begin{eqnarray}}
\newcommand{\eea}{\end{eqnarray}}

\renewcommand{\P}{\Phi(s,t)}

\newcommand{\q}{\theta}

\newcommand{\ep}{\epsilon}

\newcommand{\bD}{\bar{D}}
\newcommand{\cO}{{\cal O}}

\newcommand{\cK}{{\cal K}}

\newcommand{\cN}{{\cal N}}
\newcommand{\cZ}{{\cal Z}}
\newcommand{\vp}{\varphi}
\newcommand{\T}{\mbox{Tr}}
\renewcommand{\P}{\Phi}
\newcommand{\bP}{\bar\Phi}

\newcommand{\bt}[1]{{\bar t}}

\usepackage{amssymb,amsmath,amsfonts,epsfig,color,latexsym}
\begin{document}

%%%%%%%%%
% Front page here
\thispagestyle{empty}
%\vspace*{1cm}
\null \hfill AEI-2004-127\\
\null \hfill  LAPTH-1084/05 \\
\null \hfill ROM2F/2004/34 \\

 %\vskip-10pt \hfill {\tt hep-th/0205270}
\vskip0.2truecm
\begin{center}
\vskip 0.2truecm {\Large\bf
%\titleline
\Large{Operator mixing in $\cN=4$ SYM:\\ The Konishi anomaly revisited
 }
}\\
\vskip 1truecm
%\vfill
{\bf B. Eden$^{*}$, C. Jarczak$^{**}$,
E. Sokatchev$^{**}$ and Ya. S. Stanev$^{***}$ \\
}

\vskip 0.4truecm
%\addresses
$^{*}$ {\it Max-Planck-Institut f\"ur Gravitationsphysik,
Albert-Einstein-Institut, \\
Am M\"uhlenberg 1, D-14476 Golm, Germany}\\
\vskip .2truecm $^{**}$ {\it Laboratoire d'Annecy-le-Vieux de
Physique Th\'{e}orique  LAPTH,\\
B.P. 110,  F-74941 Annecy-le-Vieux, France\footnote{UMR 5108 associ{\'e}e {\`a}
 l'Universit{\'e} de Savoie} } \\
\vskip .2truecm $^{***}$ {\it Dipartimento di Fisica, Universit`a di Roma ``Tor Vergata"\\
I.N.F.N. - Sezione di Roma ``Tor Vergata" \\
Via della Ricerca Scientifica, 00133 Roma, Italy}\\
\end{center}

\vskip 1truecm \Large
%\noindent
\centerline{\bf Abstract} \normalsize In  the context of the superconformal $\cN=4$ SYM theory the Konishi anomaly can be viewed as the descendant $\cK_{\mathbf{10}}$ of the Konishi multiplet in the $\mathbf{10}$ of $SU(4)$, carrying the anomalous dimension of the multiplet. Another descendant $\cO_{\mathbf{10}}$ with the same quantum numbers, but this time without anomalous dimension, is obtained from the protected half-BPS operator $\cO_{\mathbf{20}'}$ (the stress-tensor multiplet). Both $\cK_{\mathbf{10}}$ and $\cO_{\mathbf{10}}$ are renormalized mixtures of the same two bare operators, one trilinear (coming from the superpotential), the other bilinear (the so-called ``quantum Konishi anomaly"). Only the operator $\cK_{\mathbf{10}}$ is allowed to appear in the right-hand side of the Konishi anomaly equation, the protected one $\cO_{\mathbf{10}}$ does not match the conformal properties of the left-hand side. Thus, in a superconformal renormalization scheme the separation into ``classical" and ``quantum" anomaly terms is not possible, and the question whether the Konishi anomaly is one-loop exact is out of context. The same treatment applies to the operators of the BMN family, for which no analogy with the traditional axial anomaly exists. We illustrate our abstract analysis of this mixing problem by an explicit calculation of the mixing matrix at level $g^4$ (``two loops") in the supersymmetric dimensional reduction scheme.

\newpage
\setcounter{page}{1}\setcounter{footnote}{0}

\section{Introduction}

Many years ago it has been realized \cite{PiguetKonishi} that the kinetic term of the $\cN=1$ chiral matter superfields $\P$, viewed as a gauge invariant composite operator (usually called the ``Konishi operator" $\cK$), satisfies an ``anomalous" conservation condition,
\begin{eqnarray}\label{001}
  \bD^2  {\cK} \equiv \bD^2 \T \left(\bP e^{gV}\P \right) &=& \T  \left(\P\frac{\pa {\cal W}(\P)}{\pa \P}\right) + \frac{g^2}{32 \pi^2} \, {\rm Tr} \left(W^\alpha W_\alpha\right) \nonumber\\
 &\equiv&  B \, + \frac{g^2}{32 \pi^2} \, F\,,
\end{eqnarray}
where ${\cal W}$ is the chiral superpotential, $V$ is the $\cN=1$ gauge superfield and $W_\alpha$ is its field strength. The first term $B$ in the right-hand side of (\ref{001}) is obtained by applying the field equations, so it is of classical origin. The second term $F$ is of purely quantum origin and is referred to as the ``quantum Konishi anomaly". Its coefficient has been obtained by a one-loop perturbative calculation.

In the free theory (no potential, no coupling to the gauge field) eq.\,(\ref{001}) defines a linear $\cN=1$ multiplet, $\bD^2 \cK = 0$.  In particular, this implies that the axial vector component $\cK = \ldots + \bar\theta\sigma^\mu\theta k_\mu(x)+\ldots$ is conserved, $\pa^\mu k_\mu=0$. This vector is sometimes called the ``Konishi current".  It should be pointed out that the conservation of the free vector does not reflect any symmetry of the interacting theory with a non-vanishing superpotential, as indicated by the classical term in (\ref{001}).\footnote{Note, however, that without the superpotential the kinetic term of the matter Lagrangian has an extra $U(1)$ symmetry and $k_\mu$ can be viewed as the corresponding axial current.} Further, the quantum term is often interpreted as an analog of the standard Adler-Bell-Jackiw axial anomaly. This analogy has been pushed even further in \cite{Grisaru,SV1}, where it is claimed that the Konishi anomaly satisfies an Adler-Bardeen theorem, i.e. its coefficient does not receive  any quantum corrections beyond one loop. This claim is substantiated by explicit two-loop perturbative calculations in \cite{Grisaru,SV1}, but in the rather special context of $\cN=1$ supersymmetric quantum electrodynamics (no matter self-interaction). The renormalization properties of the Konishi current have also been discussed in \cite{EF,AGJ}, but still without matter self-interaction. Later on, a more general statement about the one-loop exactness of the Konishi anomaly, this time for non-Abelian theories, appeared in \cite{SV2}. More recently, the same subject was discussed in \cite{CDSW} in relation to the chiral ring in supersymmetric gauge theories.

The question about the Konishi anomaly becomes particularly interesting in the context of the maximally supersymmetric ${\cN=4}$ super-Yang-Mills theory (SYM). In this theory the triplet of ${\cN=1}$ matter superfields $\P^I$ ($I=1,2,3$ is an $SU(3)$ index) are in the adjoint representations of an $SU(N)$ gauge group and have the special superpotential ${\cal W} = (g/3) \epsilon_{IJK} \T (\P^I\P^J\P^K)$. The ${\cN=4}$ SYM theory is known to be finite (i.e., its $\beta$ function vanishes). Consequently, this is a superconformal theory in four dimensions. In this context the operator $\cK$ can be viewed as a gauge invariant composite operator which gives rise to an entire ``long" ${\cN=4}$ superconformal multiplet, the so-called Konishi multiplet. It is the simplest example of an operator in the ${\cN=4}$ SYM theory having anomalous dimension.\footnote{The anomalous dimension of the Konishi multiplet has been computed at one (level $g^2$) and two (level $g^4$) loops through OPE analysis of the four-point function of stress-tensor multiplets \cite{Kon1}-\cite{Arutyunov:2000im}; recently, its three-loop value has been first predicted \cite{Beisert:2003tq} and then obtained by direct calculations \cite{Lipatov,EJS}.} It should be stressed that the Konishi operator is just the first member of the infinite family of the so-called ``BMN operators" \cite{Berenstein:2002jq}. For instance, the primary state of the dimension three BMN operator is in the \textbf{6} of $SU(4)$; in the $\cN=1$ formulation with residual R symmetry $SU(3)\times U(1)$ it is given by the superfield $\cK^I_{\mathbf{6}/3} \, = \, \T(\Phi^I \Phi^J \bar\Phi_J) + \T(\Phi^I \bar \Phi_J \Phi^J)$.\footnote{Here and in what follows the notation ${\mathbf{6}/3}$ indicates the $SU(4)$ representation and its $SU(3)$ projection.} In the free theory we find $\bD^2 \cK^I_{\mathbf{6}/3}=0$, just like in the Konishi case. In the interacting theory $\cK^I_{\mathbf{6}/3}$ obeys an ``anomalous" equation similar to (\ref{001}). The same is true for all the higher-dimensional BMN operators. However, the superspace condition $\bD^2 \cK=0$ implies the conservation of an axial current component only for the bilinear Konishi operator of dimension two. For this reason the traditional approach to the Konishi anomaly based on the analogy with the axial anomaly cannot be generalized to the higher BMN operators. Another special property of the Konishi operator which is sometimes exploited in the literature is the fact that the $B$ term in (\ref{001}) coincides with the superpotential of the $\cN=4$ theory\footnote{This point may be a source of confusion. It is known that the chiral superpotential term in the {\it action} $\int d^4x d^2\q\; {\cal W}(\Phi(x,\q))$ is subject to a non-renormalization theorem. This by no means implies that the chiral {\it operator } ${\cal W}(\Phi(x,\q))$ (i.e. the term in the Lagrangian) is protected. Indeed, a simple one-loop calculation of its two-point function shows that it is logarithmically divergent.} and the $F$ term with the $\cN=1$ SYM Lagrangian. Again, this does not generalize to the higher BMN operators. The development of a universal approach to the Konishi anomaly and to its BMN counterparts, exploiting the superconformal properties of the $\cN=4$ theory, is one of the main motivations for the present work. Another reason for it is to clarify some of the ideas of the method for calculation of the anomalous dimensions of BMN operators proposed in \cite{B} and further elaborated in \cite{EJS}.

Before describing our approach, we should recall some basic but important facts about the renormalization of composite operators (see, e.g. \cite{Collins}). In the quantum theory the operator equation (\ref{001}) should be understood as a linear relation among renormalized operators,
\begin{equation}\label{001'}
  [\bD^2  {\cK}]_R = a(g) [B]_R + b(g) [F]_R\,.
\end{equation}
Here $[\bD^2  {\cK}]_R = Z_{\cK} \bD^2  {\cK}$ is the derivative of the renormalized Konishi operator. The latter is the only scalar singlet gauge invariant operator of dimension two in the SYM theory, therefore it undergoes multiplicative renormalization $[{\cK}]_R = Z_{\cK} {\cK}$ with some divergent factor $Z_{\cK}$; the derivative $\bD^2  {\cK}$ in (\ref{001}) inherits the same renormalization factor. The operators in the right-hand side of eq.  (\ref{001'}) are the properly renormalized versions of the two terms in the right-hand side of eq.  (\ref{001}).  They are in general mixtures of the bare ones, $[B]_R = Z_{BB} B + Z_{BF} F + Z_{BK} \bD^2  {\cK}$ (and similarly for $[F]_R$), where the $Z$s form a matrix of {\it a priori} divergent renormalization factors. Finally, $a(g)$ and $b(g)$  are {\it finite} coefficients whose value depends on the normalization of the operators, i.e. on the subtraction scheme.  The standard quantum field theory prescription is that the form of the renormalized operators and the coefficients in (\ref{001'}) should be determined through insertions of the composite operators into Green's functions of elementary fields. In practice, already at two loops this procedure involves rather heavy calculations. To the best of our knowledge, such explicit calculations have been carried out in a simplified version of the model (without the superpotential term $B$) in Refs. \cite{Grisaru,SV1,EF,AGJ} using different regularization schemes. The results can be summarized as follows: if $a(g)=0$ (no superpotential), then $b(g)$ is equal to its one-loop value. The latter statement is the analog of the Adler-Bardeen theorem for this case. In the past it has been pointed out that the Adler-Bardeen theorem can be viewed as a statement about the existence of a scheme in which the anomaly coefficient is one-loop exact (see, e.g., \cite{Kelly,Ensign:1987wy} and especially \cite{Larin} where a detailed treatment of the axial anomaly up to two loops in the dimensional regularization scheme is given). Although the full renormalization procedure in the presence of the superpotential has not been explicitly worked out beyond one loop, it is generally assumed that there are no major differences and that the $F$ term in (\ref{001'}) can always be interpreted as the analog of the axial anomaly subject to the Adler-Bardeen theorem.

The main point we want to make in this paper is that the picture radically changes in the very special case of the ${\cN=4}$ SYM theory. Superconformal invariance imposes additional restrictions on the operator relation (\ref{001'}). Indeed, in the left-hand side we have an operator with well-defined conformal properties, in particular, with the anomalous dimension of the Konishi multiplet. So, the renormalized operators appearing in the right-hand side of eq.\,(\ref{001'}) must match these conformal properties. We show that there exists only one such operator, and it is the renormalized version $[B]_R$ of the ``classical anomaly" term $B$. It can be identified with the $SU(3)$ singlet projection $\cK_{\mathbf{10}/1}$ of a particular superconformal descendant $\cK_{\mathbf{10}}$ of the Konishi multiplet in the $\mathbf{10}$ of the R symmetry group $SU(4)$. It has naive dimension three but as a quantum operator it acquires the anomalous dimension of the Konishi multiplet.  On the contrary, the renormalized version $[F]_R$ of the ``quantum anomaly" term $F$ turns out to be the singlet projection $\cO_{\mathbf{10}/1} = F-4B $ of the descendant $\cO_{\mathbf{10}}$ of the so-called stress-tensor multiplet $\cO_{\mathbf{20}'}$ which has ``protected" (canonical) dimension. Our conclusion is that in the ${\cN=4}$ case the ``anomaly" equation (\ref{001'}) is truncated,
\begin{equation}\label{001''}
  [\bD^2  {\cK}]^{\cN=4}_R = a(g) [B]^{\cN=4}_R \equiv  \cK_{\mathbf{10}/1}\,.
\end{equation}
In it there simply is no room for the ``quantum anomaly" term $[F]_R$, due to the mismatch of the conformal properties. To put it differently, the bare $F$ term has been absorbed into the definition of the renormalized operator mixture $\cK_{\mathbf{10}/1}$ with a coefficient $Z_{F}$ which is in fact a divergent renormalization factor beyond one loop (such factors are related to the so-called ``matrix elements" of the operator mixture).
Exactly the same picture applies to the BMN operator of dimension three $\cK_{\mathbf{6}}$.

The idea to interpret the Konishi anomaly as a superconformal descendant of the Konishi multiplet in the framework of the ${\cN=4}$ SYM theory was proposed in \cite{I,Roma2} and was used for a practical calculation of anomalous dimensions in \cite{B,EJS}. The starting point there is the protected (also called ``short", or half-BPS, or CPO) $\cN=4$ SYM  stress-tensor supermultiplet. Its lowest (primary) component $\cO_{\mathbf{20}'}$ is a scalar of dimension two in the $\mathbf{20}'$ of the R symmetry group $SU(4)$, and the top-spin descendant of the multiplet is the conserved stress tensor. Unlike the ``long" Konishi multiplet, the short multiplet $\cO_{\mathbf{20}'}$ is protected from quantum corrections, and hence has no anomalous dimension.\footnote{The ``protectedness" of the supermultiplet $\cO_{\mathbf{20}'}$ can be explained by the presence of the conserved stress tensor among its components. However, the absence of quantum corrections to the two- and three-point functions of a whole class of BPS operators is a more general phenomenon not related to any conservation law (for reviews see \cite{HH1,FS,DF}). The absence of renormalization of the two-point functions of half-BPS operators was confirmed by explicit perturbative calculations at levels $g^2$ and $g^4$ in \cite{Penati}. } As shown in \cite{I}, applying two (non-linear on-shell) ${\cN=4}$ supersymmetry generators to the ground state $\cO_{\mathbf{20}'}$, one can construct another member (a superconformal ``descendant") of this protected multiplet, $\cO_{\mathbf{10}}$. It is a scalar of dimension three in the $\mathbf{10}$ of the R symmetry group $SU(4)$. This descendant is realized as a linear combination of two composite operators, a trilinear ($B_{\mathbf{10}}$) and a bilinear ($F_{\mathbf{10}}$) ones. In the  $\cN=1$ formulation of the theory, restricting $B_{\mathbf{10}}$, $F_{\mathbf{10}}$ to the $SU(3)$ singlet projection, one obtains the terms $B$ and $F$ which appear in the right-hand side of ``anomaly" equation (\ref{001}), but now they form a different linear combination $\cO_{\mathbf{10}/1} = -1/2(F - 4 B)$. This combination is expected to be protected, i.e. to keep its canonical dimension. Further, one could attempt to generate a similar scalar descendant $\cK_{\mathbf{10}}$ of the Konishi multiplet by applying two on-shell ${\cN=4}$ supersymmetry transformations to the operator $\cK$. In $\cN=1$ superfield language, the $SU(3)$ singlet projection $\cK_{\mathbf{10}/1}$ should be precisely the right-hand side of eq.\,(\ref{001}). However, this naive attempt fails -- one only sees the ``classical anomaly" term $B$ in (\ref{001}), but completely misses the ``quantum anomaly" $F$. The argument of \cite{I} goes on to say that the correct form (\ref{001}) of $\cK_{\mathbf{10}/1}$ can be determined by requiring orthogonality with $\cO_{\mathbf{10}/1}$, in the sense that the two-point function $\langle\bar\cK_{\mathbf{10}/1} \cO_{\mathbf{10}/1} \rangle$ must vanish. Indeed, the crucial difference between the two descendants $\cK_{\mathbf{10}}$ and $\cO_{\mathbf{10}}$ is that in the quantum theory the former acquires an anomalous dimension (that of the Konishi multiplet), while the latter is protected, hence the two operators must be orthogonal. In other words, we are dealing with a typical operator mixing problem with the additional requirements that the diagonalized operators must be eigenstates of the superconformal dilatation operator.

In this way one can indeed discover the ``missing" $F$ term in (\ref{001}), at least at one loop. The question we want to address in this paper is what happens beyond one loop. We explicitly resolve the operator mixing described above up to level $g^4$ (``two loops") in perturbation theory. We first give a general description of the expected form of the ``pure" superconformal states. Then we verify it by an explicit graph calculation, using manifestly ${\cN=1}$ supersymmetric Feynman rules and working in the supersymmetric dimensional reduction scheme (SSDR) \cite{DimRed}. We find that the protected combination $\cO_{\mathbf{10}}$ is not renormalized at all, it remains in its classical form. However, the Konishi descendant $\cK_{\mathbf{10}}$ changes its form at every loop level. At one loop we rediscover the correct coefficient of the $F$ term in (\ref{001}), but already at two loops this coefficient becomes a divergent renormalization factor. The treatment of the BMN operator $\cK_{\mathbf{6}}$ follows exactly the same lines.

On the technical side, we can profit from the superconformal properties of the $\cN=4$ theory to further simplify the problem. Generically, $[B]_R$ and $[F]_R$ are mixtures of three bare operators with the same quantum numbers, $B$, $F$ and $\bD^2\cK$. However, remembering that eq.\,(\ref{001}) is the singlet $SU(3)$ projection of a $\mathbf{10}$ of $SU(4)$, we can switch over to its projection in the $\mathbf{6}$ of $SU(3)$. The advantage is that in this channel we can obtain the protected descendant $\cO_{\mathbf{10}/6}$ directly from a suitable projection of the primary operator through superspace differentiation,  $\cO_{\mathbf{10}/6} = D^2\cO_{\mathbf{20}'/6}$. Then we can use the classical field equation to find $D^2\cO_{\mathbf{20}'/6} = -1/2 \, (F_6 - 4 B_6)$. Here we argue that this naive operator relation remains non-renormalized, unlike that for the long Konishi multiplet. This allows us to eliminate one of the operators, e.g. $F_6$. Then the Konishi descendant in this channel is obtained as a mixture of the remaining two,  $\cK_{\mathbf{10}/6} = Z_{\cK} B_6 + Z_{\cO} D^2\cO_{\mathbf{20}'/6}$. \footnote{Note that in the $\cN=1$ formulation of the $\cN=4$ theory the projection $\cK_{\mathbf{10}/6}$ cannot be obtained directly from the singlet Konishi operator $\cK$ through $\cN=1$ superspace differentiation (or, equivalently, through $\cN=1$ supersymmetry transformations). Similarly, the projection $\cO_{\mathbf{10}/1}$ cannot be obtained from the primary $\cO_{\mathbf{20}'}$ since it does not have a singlet $SU(3)$ projection.} Further, instead of determining the renormalization factors through insertions into Green's functions of elementary fields, we do so by diagonalizing the two-point functions of the descendants. According to the superconformal picture, we expect to find\footnote{Notice that these equations become exact only when the dimensional regulator is set to zero, i.e. when conformal invariance is restored.}
\begin{eqnarray}\label{diag2p}
  \langle \bar \cK_{\mathbf{10}/6} \cK_{\mathbf{10}/6}\rangle &=& \frac{C_{\cK}}{(x^2)^{3+\gamma}}  + \mbox{$\theta$ terms}\,; \nonumber\\
 \langle \bar \cO_{\mathbf{10}/6} \cO_{\mathbf{10}/6}\rangle &=& \frac{1}{(x^2)^{3}}  + \mbox{$\theta$ terms}\,; \nonumber\\
\langle \bar \cK_{\mathbf{10}/6} \cO_{\mathbf{10}/6}\rangle &=& 0\,,
\end{eqnarray}
where $\gamma(g)$ is the anomalous dimension of the Konishi multiplet.
This gives us a set of equations for determining the renormalization factors. Knowing that the primaries $\cK$ and $\cO_{\mathbf{20}'}$ (and hence their superspace derivatives) are orthogonal allows us to carry out the diagonalization in the most efficient way. Once we have found the form of the renormalized operator mixtures $\cK_{\mathbf{10}/6}$ and $\cO_{\mathbf{10}/6}$, we perform an $SU(4)$ rotation back to the singlet $SU(3)$ channel. This gives us the correct form of $\cK_{\mathbf{10}/1}$ and $\cO_{\mathbf{10}/1}$. In particular, we see that the bare operator $\bD^2\cK$ does not mix with $B$ and $F$.

The paper is organized as follows. In Section \ref{sec2} we give the detailed definitions of the descendants $\cK_{\mathbf{10}}$ and $\cO_{\mathbf{10}}$ (and of their analogs for  the BMN operator $\cK_{\mathbf{6}}$), restricting them to their most convenient $SU(3)\times U(1)$ projections. We also formulate the diagonalization problem. In Section \ref{se2} we discuss the renormalization of conformal operators in dimensional regularization. We explain how the anomalous dimension determines the poles in the renormalization factors, first in the case of a single operator (no mixing). When two operators are allowed to mix, the structure of their renormalization factors is considerably more complicated, as we show on the example of  $\cK_{\mathbf{10}}$ and $\cO_{\mathbf{10}}$. Section \ref{se3} is devoted to the graph calculation at level $g^4$.

\section{The Konishi operator $\cK$, the BMN operator $\cK_{\mathbf{6}}$, the half-BPS operators ${\cO_{\mathbf{20}'}}$ and ${\cO_{\mathbf{50}}}$ and their descendants}\label{sec2}

In this paper we study the two simplest operators from the BMN family in $\cN=4$ SYM theory, the (naive) dimension two Konishi operator $\cK$ and the dimension three operator  $\cK_{\mathbf{6}}$ and their superconformal descendants $\cK_{\mathbf{10}}$ and  $\cK_{\mathbf{45}}$, respectively. Descendants with the same quantum numbers (apart from the anomalous dimension) can also be obtained from the half-BPS operator ${\cO_{\mathbf{20}'}}$ and ${\cO_{\mathbf{50}}}$. In this section we give their description in the ${\cN=1}$ superspace formulation of the $\cN=4$ theory.

\subsection{The Konishi operator $\cK$}

The Konishi operator is just the kinetic term of the ${\cN=1}$ (anti)chiral scalar matter superfields $\Phi^I$ ($\bar\Phi_I$), $I=1,2,3$ (see the Appendix for the complete  $\cN=4$ SYM action):
\begin{equation}\label{01}
  \cK =
% \mbox{\huge ???}C_{\cK}
-\frac{1}{3}\T \left(e^{gV} \bar \Phi_I e^{-gV} \Phi^I  \right)
\end{equation}
% The constant $C_{\cK}$ is chosen to achieve the standard normalization at unity.
It is a singlet of the R symmetry group $SU(4)$ of the $\cN=4$ SYM theory, and consequently of the residual $SU(3)\times U(1) \in SU(4)$, where $SU(3)$ rotates the indices $I$ and $U(1)$ gives the matter superfields R charge, $2/3$ for $\P$ and $-2/3$ for $\bP$ in units in which the R charge of $\theta$ is $1$. In the quantum theory this operator is known to develop anomalous dimension \cite{Kon1}-\cite{EJS}.

The operator $\cK$ (\ref{01}) is an $\cN=1$ superfield, and as such it has a number of components. For instance, acting on it with two spinor derivatives $\bD^2$, we obtain a scalar of canonical dimension $3$ and of R charge $2$. Now, we can use the $\cN=1$ matter {\bf classical} field equation\footnote{The squared derivatives denote $D^2 \, = \, - \frac{1}{4} D^\alpha D_\alpha, \; \bar D^2 \, = \, - \frac{1}{4} \bar D_{\dot \alpha} \bar D^{\dot \alpha}$.}
\begin{equation}\label{06}
  \bD^2\left(e^{gV} \bP_I e^{-gV}\right) = - \frac{g}{2} \epsilon_{IJK} [\P^J,\P^K]
\end{equation}
to obtain
\begin{equation}\label{07}
   \bD^2 \cK =  \frac{g}{6} \epsilon_{IJK} {\rm Tr}\left(\P^I [\P^J,\P^K] \right) \equiv  {\cal K}_{\mathbf{10}/1}^{\rm classical}\,.
\end{equation}
Here the subscript ${\mathbf{10}/1}$ means that the operator is an $SU(3)$ singlet projection of a ${\mathbf{10}}$ of $SU(4)$ (see below).

Since the work of \cite{PiguetKonishi} we have known that in the quantum theory this equation must be corrected by an ``anomaly" term:
\begin{equation}\label{07'}
  \bD^2 \cK  =  \frac{g}{6} \epsilon_{IJK} {\rm Tr}\left(\P^I [\P^J,\P^K] \right) + \frac{g^2 N}{32 \pi^2} {\rm Tr} \left(W^\alpha W_\alpha\right) \equiv {\cal K}_{\mathbf{10}/1}^{\rm one-loop}
\end{equation}
where $W_\alpha$ is the (chiral) $N=1$ SYM field strength. The coefficient of the new term has been computed at one loop (order $g^2$).

As mentioned in the Introduction, an alternative way to view eq.\,(\ref{07'}) is as an operator mixing problem. The right-hand side of (\ref{07'}) contains two operators made out of bosonic (fermionic) superfields, hence the notation $B$ ($F$),\footnote{We find it both natural and convenient to include the factor $g$ accompanying the non-Abelian commutator $[\P^J,\P^K]$ into the definition of the operator $B$. In particular, this allows us to have a perturbative expansion in even powers of $g$.}
\begin{equation}\label{08}
  B= \frac{g}{6} \epsilon_{IJK} {\rm Tr}\left(\P^I [\P^J,\P^K] \right)\,, \qquad F={\rm Tr} \left(W^\alpha W_\alpha\right)\,,
\end{equation}
having the same quantum numbers (spin 0, dimension 3, R charge $2$, $SU(3)$ singlets). In the quantum theory such operators start mixing. We recall that $\cN=4$ SYM is a conformal theory, so the way to resolve this mixing is to find ``pure states" of the dilatation operator, i.e. mixtures which have a well-defined conformal anomalous dimension. Thus, the combination ${\cal K}_{\mathbf{10}/1}^{\rm quantum}$ which one should put in the right-hand side of (\ref{07'}) must be such that its anomalous dimension is that of the Konishi operator $\cK$. It is customary to call the operator   ${\cal K}_{\mathbf{10}/1}$ a {\it (superconformal) descendants} of the Konishi operator (or simply a member of the Konishi superconformal multiplet). So, equation (\ref{07'}) identifies two objects, the {\it superfield component} $\bD^2\cK$ with the {\it descendant} $\cK_{\mathbf{10}/1}$. We insist on this difference between component and descendant: the former is obtained by simple differentiation of the superfield, the later through use of the field equations. It is well known that in quantum field theory the naive use of the classical field equations may lead to incorrect results, and the Konishi ``anomaly" is a good example for this. Therefore, the question what is the operator realization of a particular descendant of a superconformal multiplet must be answered through quantum calculations, by resolving the corresponding mixing problem.

The full $\cN=4$ Konishi multiplet has a scalar descendant $\cK_{\mathbf{10}}$ of dimension 3 in the $\mathbf{10}$ of $SU(4)$. The $\cN=1$ descendant ${\cal K}_{\mathbf{10}/1}$ that we have derived so far is the $SU(3)\times U(1)$ projection $(\mathbf{1},2)$ in the decomposition $\mathbf{10} \ \rightarrow\ (\mathbf{1},2)+({\mathbf{3}},2/3)+(\mathbf{6},-2/3)$. The easiest way to see this\footnote{A manifestly $SU(4)$ covariant description is given in \cite{I,Roma2,EJS}.} is to imagine the generalization of the term $F$ in (\ref{08}). The $\cN=1$ SYM multiplet $W_\alpha = \lambda_\alpha(x)+\ldots$ includes one of the four gluinos  $\lambda^i_\alpha$ ($i=1,2,3,4$) of the full $\cN=4$ multiplet. The $SU(4)$ covariant counterpart of (the first component of) $F$ then is $\lambda^i_\alpha\ep^{\alpha\beta}\lambda^j_\beta = \lambda^j_\alpha\ep^{\alpha\beta}\lambda^i_\beta$, which indeed forms a $\mathbf{10}$ of $SU(4)$. In the $\cN=1$ formulation the other three gluinos are contained in the matter sector, $D_\alpha \P^I = \lambda_\alpha^I(x)+\ldots$. Thus, instead of $F$ (\ref{08}), we can study a different projection of the $\mathbf{10}$, for example,
\begin{equation}\label{09}
  F^{IJ}_6 = F^{JI}_6 = C_F{\rm Tr} \left(\nabla^\alpha \P^I \nabla_\alpha \P^J\right)\,,\qquad \nabla_\alpha \P^I = e^{gV} \left[D_\alpha\left(e^{-gV} \P^I e^{gV}  \right) \right] e^{-gV}\,.
\end{equation}
% {\bf\Large What is $C_F$ if $F$ is normalized at unity?}
This is an operator in  the $(\mathbf{6},-2/3)$ of $SU(3)\times U(1)$.
Similarly, the operator $B$ in (\ref{08}) is part of a $\mathbf{10}$ of $SU(4)$; its counterpart
\begin{equation}\label{010}
  B^{IJ}_6 = \frac{g}{4}\ep^{IKL}{\rm Tr} \left(\P^J [\bP_K,\bP_L]\right) + (I \leftrightarrow J)
\end{equation}
is in the same $SU(3)\times U(1)$ representation as $F^{IJ}_6$.
% {\bf\Large What is $C_B$ if $B$ is normalized at unity?}
Together they can form a mixture which is the counterpart of $\cK_{\mathbf{10}/1}$ (\ref{07'}):
\begin{equation}\label{011}
  \cK^{IJ}_{\mathbf{10}/6} = Z_{B} B^{IJ}_6 + Z_{F}F^{IJ}_6\,.
\end{equation}
We have denoted the mixing coefficients in (\ref{011}) $Z_B,Z_F$ in anticipation of their nature of renormalization factors (see Section \ref{se2}). In Section \ref{opid} we will show that at the lowest order ($g^2$) of perturbation theory they have the same values as in the mixture (\ref{07'}), $Z_B=1$, $Z_F = \frac{g^2 N}{32 \pi^2}$. This is a consequence of the fact that $\cK_{\mathbf{10}/1}$ and $\cK_{\mathbf{10}/6}$ belong to the same $SU(4)$ multiplet $\cK_{\mathbf{10}}$, and in the $\cN=4$ theory the $SU(4)$ symmetry should be exact at the quantum level. The quantum calculations at order $g^2$ are very simple, no matter whether we study the projections $\cK_{\mathbf{10}/1}$ or $\cK_{\mathbf{10}/6}$ of $\cK_{\mathbf{10}}$, but at the next level $g^4$ it is much more convenient to work with $\cK_{\mathbf{10}/6}$.

We remark that the complete $\cN=4$ descendant $\cK_{\mathbf{10}}$ of the Konishi multiplet involves a third projection $({\mathbf{3}},2/3)$, i.e. the scalar operator $\cK^{I}_{\mathbf{10}/3} = \frac{g}{2} {\rm Tr} \left(\P^J [\bP_J,\P^I]\right) + \frac{g^2 N}{32 \pi^2} {\rm Tr} \left(\nabla^\alpha \P^I W_\alpha\right)$. Among the three projections $\cK_{\mathbf{10}/1}$, $\cK^{I}_{\mathbf{10}/3}$ and $\cK^{IJ}_{\mathbf{10}/6}$ of $\cK_{\mathbf{10}}$, only $\cK_{\mathbf{10}/1}$ is a descendant of $\cK$ in the restricted $\cN=1$ sense, i.e. only it can be obtained directly through the (anomalous) use of the $\cN=1$ field equations.

Finally, we recall that in the quantum theory the Konishi multiplet acquires an anomalous dimension.\footnote{The terms ``anomalous dimension" and ``Konishi anomaly" should not be confused. The former is a general property of all long (unprotected) superconformal multiplets in the $\cN=4$ theory. The latter, as we argue here, evokes an analogy with the standard axial anomaly which is somewhat misleading. } This means that its renormalized two-point function has the form
\begin{equation}\label{011'}
  \langle \cK \cK\rangle = \frac{1}{(x^2)^{2+\gamma}}  + \mbox{$\theta$ terms}\,,
\end{equation}
where $\gamma(g^2)= \gamma_1 g^2 + \gamma_2 g^4 + \ldots$ is given by a perturbative expansion. The descendants $\cK_{\mathbf{10}/1}$ and $\cK_{\mathbf{10}/6}$ are supposed to belong to the same superconformal multiplet, so they must have the same anomalous dimension $\gamma$.

\subsection{The half-BPS operator ${\cO_{\mathbf{20}'}}$}

The two operators $F_6$ (\ref{09}) and $B_6$ (\ref{010}) can form another mixture orthogonal to $\cK_{\mathbf{10}/6}$. This is a descendant of the other $\cN=4$ scalar multiplet of dimension two, the half-BPS operator ${\cO_{\mathbf{20}'}}$. The latter has its primary state (lowest component) in the $\mathbf{20}'$ of $SU(4)$ whose decomposition under $SU(3)\times U(1)$ is $\mathbf{  20}' \ \rightarrow\ (\mathbf{  6},4/3)+(\bar{\mathbf{  6}},-4/3)+(\mathbf{  8},0)$. In terms of $\cN=1$ superfields the projection $(\mathbf{  6},4/3)$ is realized by the chiral superfield
\begin{equation}\label{012}
  {\cO}^{IJ}_{\mathbf{20}'/6} = C_{\cO}  {\rm Tr}(\Phi^I\Phi^J)\,,
\end{equation}
where $C_{\cO}$ is a normalization constant. The operator ${\cO_{\mathbf{20}'}}$ has the remarkable property of being ``protected" from quantum corrections, i.e., its two-point function keeps its tree-level form (the $SU(4)$ indices are suppressed):
\begin{equation}\label{013}
  \langle \bar\cO_{\mathbf{20}'}(1) \cO_{\mathbf{20}'}(2)\rangle = \frac{1}{(x^2_{12})^2} + \mbox{$\theta$ terms}\,.
\end{equation}
In particular, this means that it has no anomalous dimension.
The fact that the projection ${\cO}_{\mathbf{20}'/6}$ (\ref{012}) of the operator ${\cO_{\mathbf{20}'}}$ is chiral and also primary (in the sense that it cannot be obtained from any other operator of lower dimension through use of the field equations) has given rise to the popular term ``chiral primary operator" (CPO).\footnote{A superconformal primary operator satisfying a BPS shortening condition must have a fixed, ``quantized" dimension (for reviews see \cite{HH1,FS}). Whether a given composite operator is primary or not is a subtle question which can only be fully answered in the quantum theory. Some indications how to recognize ``Chiral Primary Operators" were given in \cite{Witten}, and a more elaborate criterium was proposed in \cite{HH2}.}   We prefer to use the more relevant term ``half-BPS operator". Indeed, the half-BPS multiplet $\cO_{\mathbf{20}'}$ also has the $SU(3)\times U(1)$ projection $(\mathbf{  8},0)$,
\begin{equation}\label{014}
  \left({{\cO}_{\mathbf{20}'/8}}\right)_I{}^J = {\rm Tr}\left(e^{gV}\bP_I e^{-gV}\P^J\right) - \frac{1}{3}\delta_I^J\;{\rm Tr}\left(e^{gV}\bP_K e^{-gV}\P^K\right)\,,
\end{equation}
which is not chiral but is nevertheless protected.

The chiral superfield (\ref{012}) is short, the top component in its $\theta$ expansion, $D^2{\cO}_{\mathbf{20}'/6}(\q=0)$, is a scalar of dimension three  in the $(\mathbf{6},-2/3)$ of $SU(3)\times U(1)$. Just as we did with the Konishi operator $\cK$, we can obtain a descendant of this protected multiplet by using the classical field equations:
\begin{equation}\label{015}
   D^2{\cO}^{IJ}_{\mathbf{20}'/6} \, = \, - \frac{1}{2}(F^{IJ}_6 \, - \, 4 B^{IJ}_6) \,\equiv \, \cO^{IJ}_{\mathbf{10}/6} \,,
\end{equation}
where $F,B$ have been defined in (\ref{09}), (\ref{010}). Note the important difference between $\cK_{\mathbf{10}/6}$ and $\cO_{\mathbf{10}/6}$: The former is a descendant of the Konishi multiplet only in the $\cN=4$ sense, while the latter is a descendant of the half-BPS multiplet also in the $\cN=1$ sense.

In principle, in the quantum theory the coefficients in (\ref{015}) may get renormalized,
\begin{equation}\label{016}
  D^2{\cO}^{IJ}_{\mathbf{20}'/6} = \cZ_{F}F^{IJ}_6 + \cZ_{B} B^{IJ}_6\equiv \cO^{IJ}_{\mathbf{10}/6}\,.
\end{equation}
One of the aims of this paper is to show that in the renormalization scheme we use eq.\,(\ref{015}) remains exact; in Section \ref{opid} we verify this up to $g^2$ in correlation functions with respect to $D^2 {\cal O}$ and $F$, and up $g^4$ with respect to $B$. In a sense, eq.\,(\ref{015}) is natural for an operator mixture which is protected, i.e. whose two-point function keeps its tree-level form,
\begin{equation}\label{017}
  \langle \bar \cO_{\mathbf{10}/6} \cO_{\mathbf{10}/6}\rangle = \frac{1}{(x^2)^3} + \mbox{$\theta$ terms}\,.
\end{equation}
In other words, we would guess that in this case the naive use of the classical field equations is justified. However, we are not aware of any general field theory criterium which would allow us to tell when the classical field equations do receive corrections (the Konishi operator and its descendants are an example) and when they do not.\footnote{This may be a scheme-dependent property.}  Therefore we find it necessary to carry out an explicit quantum calculation in a particular scheme, which confirms our conjecture that the mixture (\ref{015}) is not renormalized while the other one, (\ref{011}), receives quantum corrections. The fact that eq.\,(\ref{015}) remains exact at the quantum level is very useful, it allows us to eliminate one of the three operators, e.g. $F_6$ (see Section \ref{1.3}).

\subsection{Diagonalization of the operators $B$ and $F$}\label{1.3}

Let us summarize the discussion so far. In the $\cN=1$ formulation of the $\cN=4$ SYM theory there exist two gauge invariant scalar composite operators in the $(\mathbf{  6},-2/3)$ of $SU(3)\times U(1)$, $F_6$ (\ref{09}) and $B_6$ (\ref{010}). From them we can prepare two independent mixtures. The first is the descendant $\cK_{\mathbf{10}/6}$ of the Konishi multiplet and as such it must have the same anomalous dimension $\gamma$. In other words, its two-point function should have the form
\begin{equation}\label{018}
  \langle \bar \cK_{\mathbf{10}/6} \cK_{\mathbf{10}/6}\rangle = \frac{C_{\cK}}{(x^2)^{3+\gamma}}  + \mbox{$\theta$ terms}\,.
\end{equation}
This follows from the fact that $\cK_{\mathbf{10}/6}$ and $\cK$ belong to a superconformal multiplet of operators which carry different canonical dimensions, but the same anomalous dimension. This operator is a projection of the $\mathbf{10}$ of $SU(4)$ which corresponds to the complete $\cN=4$ superdescendant $\cK_{\mathbf{10}}$ of of the Konishi multiplet. Note that within the $\cN=1$ framework this particular representative of $\cK_{\mathbf{10}}$ cannot be obtained directly from $\cK$ through the field equations, unlike the $SU(3)$ singlet $\cK_{\mathbf{10}/1}$ (\ref{07'}). The relationship between $\cK_{\mathbf{10}/6}$ and $\cK_{\mathbf{10}/1}$ is indirect, it evokes the full $\cN=4$ supersymmetry of the theory which is not manifest in the $\cN=1$ formulation. The situation changes in the harmonic superspace formulation with manifest ${\cN=2}$ supersymmetry \cite{HSS}. There one can obtain \cite{AS} a direct descendant of $\cK$ which involves two of the four gluinos of the $\cN=4$ theory, and thus mixes together the fermion operators $F$ (\ref{08}) and $F_6$ (\ref{09}). The same applies to the operators $B$ and $B_6$. The existence of such a formulation with a larger manifest supersymmetry is an additional justification of the path we have chosen to follow here. We want to resolve the mixing problem of the operators $F_6$ and $B_6$ and then to use the same mixing matrix for $F$ and $B$.

The other mixture is the superdescendant $\cO_{\mathbf{10}/6}$ of the protected half-BPS operator $\cO_{\mathbf{20}'/6}$. Its two-point function is given in (\ref{017}), i.e. it has no anomalous dimension. Since both operators $\cK_{\mathbf{10}/6}$ and $\cO_{\mathbf{10}/6}$ are supposed to be pure conformal states of the same canonical, but different anomalous dimension, they must be orthogonal to each other,
\begin{equation}\label{019}
  \langle \bar\cO_{\mathbf{10}/6} \cK_{\mathbf{10}/6}\rangle = 0\,.
\end{equation}

The {\it superdescendant} $\cO_{\mathbf{10}/6}$ is related to the {\it supercomponent} $D^2\cO_{\mathbf{20}'/6}$ via the (quantum) dynamical equation (\ref{016}). This fact can be used to make a change of basis in the operator mixture $\cK_{\mathbf{10}/6}$.\footnote{Before doing this it is important to make sure that the $Z$ factors in (\ref{016}) are not divergent. The reason is that equations like, e.g. the orthogonality condition (\ref{019}) only hold up to  $O(\epsilon)$ terms where  $\epsilon$ is the dimensional regulator (they only become exact in the limit $\epsilon\to 0$). So, dividing by singular $Z$ factors may create uncontrollable finite contributions (see Section \ref{se2} for details). In the case of eq.\,(\ref{016}) not only the $Z$ factors are finite, but they are given by the classical expression (\ref{015}), as we argue in this paper.  } For example, we can eliminate the operator $F_6$ in favor of $D^2\cO_{\mathbf{20}'/6}$ and replace (\ref{011}) by
\begin{equation}\label{020}
  \cK_{\mathbf{10}/6} = Z_{\cK} B_6 + Z_{\cO}D^2\cO_{\mathbf{20}'/6}\,.
\end{equation}
The mixing (renormalization) factors $Z_{\cK},\ Z_{\cO}$ are of course different from those in (\ref{011}). The notation $Z_{\cK}$ for the factor of $B_6$ suggests that it will turn out identical (up to a finite overall normalization) with the renormalization factor of the primary Konishi operator $\cK$ (see Section \ref{s3}). Similarly, the orthogonality condition (\ref{019}) can be equivalently rewritten as
\begin{equation}\label{021}
  \langle \bD^2\bar \cO_{\mathbf{20}'/6} \cK_{\mathbf{10}/6}\rangle = 0\,.
\end{equation}
Finally, the analog of (\ref{017}) becomes
\begin{equation}\label{022}
  \langle \bD^2\bar \cO_{\mathbf{20}'/6} D^2 \cO_{\mathbf{20}'/6}\rangle = \frac{1}{(x^2)^3} + \mbox{$\theta$ terms}\,.
\end{equation}
It should be stressed that eq.\,(\ref{022}) is an obvious consequence (just a derivative) of equation (\ref{013}) stating that the operator $\cO_{\mathbf{20}'/6}$ is a member of a protected multiplet. On the other hand, eq.\,(\ref{017}) is a non-trivial condition on the operator mixture $\cO_{\mathbf{10}/6}$. The difference comes from the dynamical nature of the relation (\ref{016}).

In the rest of the paper we shall see that the abstract analysis of the mixing problem, as well as the actual graph calculations are considerably simplified if we use the form (\ref{020}) instead of (\ref{011}). The reason for this is that while studying the mixture $\cK_{\mathbf{10}/6}$ we can treat $D^2 \cO_{\mathbf{20}'/6}$ as a pure state, thus avoiding the simultaneous determination of the two mixtures $\cK_{\mathbf{10}/6}$ and $\cO_{\mathbf{10}/6}$.

\subsection{The BMN operator $\cK_{\mathbf{6}}$, the short operator $\cO_{\mathbf{50}}$ and their descendants}

In ${\cal N}=1$ notation the primary operator in the long BMN
multiplet of dimension 3 is
\begin{equation}
{\cal K}_{\mathbf{6}/3}^I \, = \, \T (\Phi^I \Phi^J \bar \Phi_J) +
\T (\Phi^I \bar \Phi_J \Phi^J)
\end{equation}
and it has a higher component $\bar D^2 {\cal
K}_{\mathbf{6}/3}^I$, which may mix with
\begin{eqnarray}
B_{\mathbf{45}/3}^I & = & \frac{g}{4} \epsilon_{JKL} \, \T (\Phi^I \Phi^J
[\Phi^K, \Phi^L]) \,
, \\ F_{\mathbf{45}/3}^I & = & \T (\Phi^I W^\alpha W_\alpha) \, .
\end{eqnarray}
Exactly as in the case of the Konishi multiplet, the field equation
of the antichiral superfield implies the existence of an operator relation
\begin{equation}
Z_K \, \bar D^2 {\cal K}_{\mathbf{6}/3}^I \, + \, Z_B \, B_{\mathbf{45}/3}^I
\, + \, Z_F \, F_{\mathbf{45}/3}^I \, = \, [\bar D^2 {\cal K}_{
\mathbf{6}/3}^I]_R \, + \, a(g^2) \, [B^I]_R \,
\end{equation}
with a finite coefficient function $a(g^2)$. The existence of such a linear relation
allows us to eliminate $\bar D^2 {\cal K}_{\mathbf{6}/3}^I$ from the mixing
problem in the $\mathbf{3}$ of $SU(3)$. It is quite cumbersome to obtain the finite
constant of proportionality $a(g^2)$ in a direct calculation. As before, we
sidestep the problem by appealing to $SU(4)$: we assert that the
renormalization factors $Z_B$ and $Z_F$ are identical in all components of
the \textbf{45} of $SU(4)$ and fix them by orthogonalization
in the $\mathbf{10}$ of $SU(3)$. Orthogonalization does not permit to fix
the finite constant of proportionality $a(g^2)$, which we leave undetermined.
Hence we do not work out the anomaly equation itself. On the other hand, we
gain the freedom of choosing $Z_B$ in minimal subtraction form.

The appearance of the bare operator $\bar D^2 {\cal K}_{\mathbf{6}/3}^I$ in
$[B]_R$ is not expected, since there is no such operator in the $\mathbf{10}$ of
$SU(3)$. Although such an effect was observed e.g. in \cite{Larin} it cannot
occur in this situation because of the underlying $SU(4)$ symmetry.

The protected multiplet with a descendant in the $\mathbf{45}$ of $SU(4)$ is
\begin{equation}
{\cal O}_{\mathbf{50}/10}^{(IJK)} \, = \, \T (\Phi^{(I} \Phi^J
\Phi^{K)}) \,.
\end{equation}
It has a higher component $D^2 {\cal O}_{\mathbf{50}/10}$ which can
mix with
\begin{eqnarray}
B_{\mathbf{45}/10}^{(IJK)} & = & \frac{g}{2} \epsilon^{LM(I} \, \T (\Phi^J
\Phi^{K)} [\bar \Phi_L, \bar \Phi_M]) \, , \\
F_{\mathbf{45}/10}^{(IJK)} & = & \T (\Phi^{(I} \nabla^\alpha \Phi^J
\nabla_\alpha \Phi^{K)}) \, .
\end{eqnarray}
The field equation of the chiral superfield suggests the operator relation
\begin{equation}
{\cal O}_{\mathbf{50}/10}^{(IJK)} \, + \, \frac{3}{2} \, (F_{\mathbf{45}/10}^{(IJK)} \, - \, 2  \, B_{\mathbf{45}/10}^{(IJK)}) \, = \, 0 \, , \label{op45id}
\end{equation}
which we have checked by the same means (and to the same order) as for
${\cal O}_{\mathbf{20'}/6}$, i.e. by doing the $D$-algebra for the supergraphs
and exploiting partial integration to reduce to a set of basic $x$-space
integrals. The calculation is nearly the same as the one in Section
\ref{opid}; some new diagrams can be drawn, but they vanish. The
superposition of the other supergraphs can be simplified by exactly the same
manipulations as for ${\cal O}_{\mathbf{20'}/6}$. Once again, to the given
order eq.\,(\ref{op45id}) is true without the need of introducing
renormalization factors.
%Further, it appears to be an identity: the right-hand side is exactly zero.

\section{Dimensional regularization and anomalous dimension}\label{se2}

\subsection{General structure of the two-point functions}\label{2.1}

The quantum calculation we plan to carry out will be done in the scheme of dimensional regularization (or, more precisely, supersymmetric dimensional reduction (SSDR) \cite{DimRed}, a scheme which preserves manifest supersymmetry).  In dimensional regularization the action of a (scalar) field takes  the form
\begin{equation}\label{1}
  S =  \int d^{4-2\ep}x\; {\cal L}(\phi(x),\hat g)\,, \qquad \hat g = g\mu^{\ep}\,.
\end{equation}
Here $\mu$ is a mass parameter which allows the coupling $g$ to remain dimensionless. On the contrary, the fields $\phi$ change their dimension. For example, the propagator for a (scalar) field becomes
 \begin{equation}\label{2}
  \langle\bar\phi\phi\rangle \sim \frac{1}{(x^2)^{1-\ep}}\,.
\end{equation}
We can say that even the free field acquires a small ``anomalous dimension" $-\ep$. Of course, in the limit $\ep\to0$ this anomalous dimension disappears. This is a peculiarity of the dimensional regularization scheme.

We are interested in the two-point functions of composite operators. At each level of perturbation theory they have a general structure which is explained below. Take, for instance, the two-point function of the bilinear Konishi operator $\cK \sim \bar\phi\phi$ at tree level ($g^0$). It is described by a one-loop (in the standard, momentum space counting) graph without interaction vertices. Its $x$-space expression simply is the square of (\ref{2}):
\begin{equation}\label{2'}
  \langle {\cal K}{\cal K}\rangle_{g^0} \sim \frac{1}{(x^2)^{2-2\ep}}\,.
\end{equation}
At the first non-trivial level $g^2$ the graphs have two loops and two interaction vertices, etc. In general, an $n$-loop two-point function graph for $\cK$ has the dimensionful factor
\begin{equation}\label{4}
  \langle {\cal K}{\cal K}\rangle_{n\ {\rm loop}} \ \Rightarrow\  (g^2\mu^{2\ep})^{n-1}\,.
\end{equation}
Similarly, for an operator made out of $m$ fields (for $\cK$ $m=2$), whose free (order $g^0$ or, abusing the term, ``tree-level") two-point function has $m-1$ loops, we find  \begin{equation}\label{5}
  \langle \bar{\cal O}^m {\cal O}^m\rangle_{n\ {\rm loop}}  \ \Rightarrow\  (g^2\mu^{2\ep})^{n-m+1}\,.
\end{equation}
The important point here is that the perturbative expansion goes in powers of the single variable $\hat g^2 = g^2\mu^{2\ep}$.

Thus, the general structure of an $n$-loop two-point function graph for the $m$-linear operator ${\cal O}^m$ is
\begin{eqnarray}
   \langle \bar{\cal O}^m {\cal O}^m\rangle_{n\ {\rm loop}}  &=& \left(\frac{c_{n-m+1,n-m+1}}{\ep^{n-m+1}} + \cdots +  \frac{c_{n-m+1,1}}{\ep} + c_{n-m+1,0} + O(\ep) \right) \nonumber\\
  &\times&\; \frac{1}{(x^2)^{m(1-\ep)}}\; [g^2(x^2\mu^2)^\ep]^{n-m+1}\;. \label{6}
\end{eqnarray}
Here the dependence on $x$ is determined by the already known dependence on $\mu$ and by the requirement that the two-point function as a whole must keep the ``engineering"  dimension $m(1-\ep)$ of the operator ${\cal O}^m$. The poles in the regulator $\ep$ come from the expansion of the divergent $n$-loop integrals of the corresponding graphs. The fact that the leading singularity in (\ref{6}) has the same order as the power of $g^2$ has to do with the renormalizability of the operator, see the next subsection.

The conclusion from the above discussion is that the perturbative two-point function of the naked operator ${\cal O}^m$ can be viewed as a function of two variables,
\begin{equation}\label{7}
  (x^2)^m\langle \bar{\cal O}^m {\cal O}^m\rangle = (x^2)^{m\ep} G(g^2(x^2\mu^2)^\ep,\ep)\,.
\end{equation}
The dimensionful parameter $\mu$ in (\ref{7}) is not essential for our subsequent analysis and can be suppressed. If needed, it can easily be restored by simple dimension counting.

\subsection{Renormalization of a single operator}

The ${\cN=4}$ SYM theory is supposed to be conformal, i.e. it has vanishing $\beta$ function. This means that the coupling $g$ is not renormalized. At the same time,  composite operators have inherent divergences which are responsible for their anomalous dimension. To be more specific, let us consider the Konishi operator $\cK$. Its advantage is its low dimension, so it is a ``pure" state, i.e. cannot mix with any other operator in the $\cN=4$ theory.   Thus, we expect that after the multiplicative renormalization of $\cK$ its two-point function (\ref{7}) will take the following form:
\begin{equation}\label{8}
  (x^2)^2\langle [\cK]_R [\cK]_R\rangle = Z_\cK^2(g^2,\ep) (x^2)^{2\ep} G(g^2x^{2\ep},\ep) = C(g^2)(x^2)^{-\gamma(g^2)} + O(\ep)
\end{equation}
(we have dropped $\mu$). Here $\gamma(g^2) = \gamma_1 g^2 + \gamma_2 g^4 + \ldots$ is the anomalous dimension of the renormalized operator $[\cK]_R = Z_\cK{\cK}$ and $Z_\cK(g^2,\ep)$ is a {\it constant} renormalization factor. The r\^ole of $Z_\cK$ is to remove all the singularities from $G$, so that the left-hand side of (\ref{8}) becomes finite. After that we can take the limit $\ep\to0$, and {\it only in this limit} we expect to find the conformal power behaviour $(x^2)^{-\gamma(g^2)}$. It is important to realize that the regulator $\ep$ (or, equivalently, the presence of the mass parameter $\mu$) breaks conformal invariance, so the $O(\ep)$ terms in the right-hand side of (\ref{8}) form a complicated function of $x$. It is clear that the factor $Z_\cK$ is determined up to an overall finite renormalization factor which modifies the normalization $C(g^2)$ of the two-point function (\ref{8}).

Now, let us take the log of eq.\,(\ref{8}):
\begin{equation}\label{9}
 2\ln Z_\cK + \ln G = -\gamma \ln x^2 + \ln C + O(\ep)\,.
\end{equation}
Further, let us differentiate (\ref{9}) with $\pa/\pa g^2$:
\begin{equation}\label{10}
  2 \frac{\pa}{\pa g^2} \ln Z_\cK + x^{2\ep} \frac{\pa }{\pa g^2x^{2\ep}}\ln G = -\frac{\pa\gamma}{\pa g^2} \ln x^2 + \frac{\pa C}{\pa g^2} + O(\ep)\,.
\end{equation}
Next, let us take the derivative $x^2\pa/\pa x^2$ of (\ref{9}):
\begin{equation}\label{11}
  \ep g^2 x^{2\ep} \frac{\pa }{\pa g^2x^{2\ep}}\ln G = -\gamma + O(\ep)\ \Rightarrow \ x^{2\ep} \frac{\pa }{\pa g^2x^{2\ep}}\ln G =  -\frac{\gamma}{\ep g^2} + O(1)\,.
\end{equation}
With the help of (\ref{11}) we can rewrite (\ref{10}) as follows:
\begin{equation}\label{12}
  2 \frac{\pa}{\pa g^2} \ln Z_\cK - \frac{\gamma}{\ep g^2} = -\frac{\pa\gamma}{\pa g^2} \ln x^2 + \frac{\pa C}{\pa g^2} + O(1)\,.
\end{equation}
Note that the left-hand side of (\ref{12}) does not depend on $x$, so the $O(1)$ terms in  the right-hand side must compensate the $\ln x^2$ term.

It is clear that the differential equation (\ref{12}) only determines the pole structure of $Z_\cK$, its $O(1)$ part is kept arbitrary. We can use this freedom in order to choose $Z_\cK$ such that $Z_\cK(0,\ep)=1$ and that it contains only singular terms. This is the so-called ``minimal subtraction" (MS) renormalization scheme. Then the solution to eq.\,(\ref{12}) is
\begin{equation}\label{13}
  Z_\cK(g^2,\ep) = \exp\left\{\frac{1}{2\ep}\int^{g^2}_0 \frac{\gamma(\tau)}{\tau}d\tau\right\} = 1 + \frac{\gamma_1}{2\ep} g^2 + \left(\frac{\gamma^2_1}{8\ep^2} + \frac{\gamma_2}{4\ep}  \right)g^4 + O(g^6)\ .
\end{equation}
Looking at the result (\ref{13}), we can easily explain the general structure of the naked two-point function $G$ (\ref{6}). Indeed, the leading singularity at the level $g^{2k}$ in the expansion of (\ref{13}) is $\sim  g^{2k}\ep^{-k}\gamma^k_1$. This term can cancel the analogous pole in the expansion of $G$ provided that the leading pole at the level $g^{2k}$ in (\ref{6}) is also of order $\ep^{-k}$. So, the form (\ref{6}) ensures the  renormalizability of the composite operators $\cO$.

Now we can restore the $\mu$ dependence. In the presence of $\mu$ eq.\,(\ref{8}) reads
\begin{equation}\label{15}
  Z_\cK^2(g^2,\ep)G(g^2(x^2\mu^2)^{\ep},\ep) = C(g^2)(x^2\mu^2)^{-\gamma(g^2)} + O(\ep)\,.
\end{equation}
The conformal renormalized operator $[\cK]_R$ should have the two-point function
\begin{equation}\label{16}
  \lim_{\ep\to0} (x^2)^2\langle [\cK]_R [\cK]_R\rangle = C(g^2)(x^2)^{-\gamma(g^2)}\,,
\end{equation}
where $\gamma$ is identified with the anomalous dimension. This is achieved by absorbing the $\mu$ dependence into the renormalization factor:
\begin{equation}\label{17}
  \hat Z_\cK(g^2,\ep; \mu) = \mu^{\gamma(g^2)} Z_\cK(g^2,\ep)\,.
\end{equation}

Finally, let us make the connection with the renormalization group (or Callan-Symanzik) equation. Remembering that the mass parameter $\mu$ can be associated the coupling constant, $\hat g = g \mu^\ep$, we can rewrite eq.\,(\ref{13}) as follows:
\begin{equation}\label{14}
  \ln Z_\cK(\hat g^2,\ep)  = \frac{1}{2\ep}\int^{\hat g^2}_0 \frac{\gamma(\tau)}{\tau}d\tau \ \Rightarrow\ \mu \frac{\pa}{\pa\mu} \ln Z_\cK = \gamma(\hat g^2) \,.
\end{equation}
We can say that this is the renormalization group equation in a conformal theory (i.e., with vanishing $\beta$ function). A peculiarity of the dimensional regularization scheme is that the $Z$ factors depend on the dimensionful ``coupling" $\hat g$. This dependence can be factored out as shown in (\ref{17}), after which the renormalization group equation takes the form (\ref{12}), where the derivatives are taken with respect to $g^2$ rather than $\mu$.

\subsection{Renormalization and mixing in the case of $\cK_{\mathbf{10}/6}$ and $D^2\cO_{\mathbf{20}'/6}$}\label{s3}

The above procedure can be adapted to the case of several operators which mix among themselves. Here we do this in the simplest case of two operators, one of which has anomalous dimension but the other is already a pure superconformal state and is protected, i.e., it has vanishing anomalous dimension. The former is the Konishi descendant $\cK_{\mathbf{10}/6}$ in the form (\ref{020}), the latter is the component $D^2\cO_{\mathbf{20}'/6}$ of the half-BPS operator $\cO_{\mathbf{20}'/6}$. In Section \ref{1.3} we explained that these operators should satisfy two conditions, (\ref{018}) and (\ref{021}) (condition (\ref{022}) is a trivial consequence of the fact that the operator $\cO_{\mathbf{20}'/6}$ is protected). Let us now see how all this works in the quantum theory. We need to know the two-point functions of the two bare operators $B_6$ and $D^2\cO_{\mathbf{20}'/6}$ in the mixture (\ref{020}). From the Feynman rules and from the discussion in Section \ref{2.1} we can derive the following general structure
\begin{eqnarray}
  \langle \bar B_6B_6\rangle &=&  g^2 (x^2)^{-3+3\ep}[1 + g^2(x^2)^{\ep} a_1(\ep)  + g^4(x^2)^{2\ep} a_2(\ep) + \ldots] \nonumber\\
  \langle \bar B_6 D^2 \cO_{\mathbf{20}'/6}\rangle = \langle \bD^2\bar \cO_{\mathbf{20}'/6} B_6\rangle &=&  g^2(x^2)^{-3+3\ep}[b_0(\ep) + g^2(x^2)^{\ep} b_1(\ep)+ \ldots]   \label{31}\\
  \langle \bD^2\bar \cO_{\mathbf{20}'/6} D^2 \cO_{\mathbf{20}'/6}\rangle &=&  (x^2)^{-3+2\ep}[1 + g^2(x^2)^{\ep} c_1(\ep)  + g^4(x^2)^{2\ep} c_2(\ep) + \ldots]\,.    \nonumber
\end{eqnarray}
Here the coefficients $a_i(\ep)$ and $b_i(\ep)$ involve poles following the general pattern (\ref{6}). The coefficients $c_i(\ep) \sim O(\ep)$ are such that in the limit $\ep\to0$ they give rise to contact terms in the two-point function of the protected operator \cite{Penati}. The expansion of $\langle \bar B_6B_6\rangle$ and $\langle \bD^2\bar \cO_{\mathbf{20}'/6} D^2 \cO_{\mathbf{20}'/6}\rangle$ start with unity, which amounts to tree-level normalization. The overall factor $g^2$ in $\langle \bar B_6 D^2 \cO_{\mathbf{20}'/6}\rangle$ is due to the fact that the first non-trivial graph involves one chiral matter coupling.

We can organize the above two-point functions into a $2\times 2$ matrix:
\begin{equation}\label{32}
     \begin{pmatrix}
    \langle \bar B_6B_6\rangle & \langle \bar B_6 D^2 \cO_{\mathbf{20}'/6}\rangle \\
    \langle \bD^2\bar \cO_{\mathbf{20}'/6} B_6\rangle & \langle \bD^2\bar \cO_{\mathbf{20}'/6} D^2 \cO_{\mathbf{20}'/6}\rangle
  \end{pmatrix}  \equiv (x^2)^{-3+2\ep} G(g^2x^{2\ep},\ep)\,,
\end{equation}
where $G$ is the matrix analog of the function in (\ref{7}) (we drop the dimensionful constant $\mu$). This time renormalization means to bring this matrix into diagonal form where we could read off (in the limit $\ep\to0$) the anomalous dimension $\gamma(g^2)$ of the Konishi multiplet (we recall that $\cK_{\mathbf{10}/6}$ is a descendant of $\cK$ and so must have the same anomalous dimension) and  the vanishing anomalous dimension of the protected operator $\cO_{\mathbf{10}/6}$. This can be achieved with the help of a constant singular renormalization (or mixing) matrix:
\begin{equation}\label{33}
  Z(g^2,\ep)G(g^2x^{2\ep},\ep)Z^\dagger(g^2,\ep) =
  \begin{pmatrix}
    C_{\cK}(g^2) (x^2)^{-\gamma(g^2)} & 0 \\
    0 & 1
  \end{pmatrix} + O(\ep)\,.
\end{equation}
As in the case of a single operator, the conformal behavior indicated in the right-hand side of (\ref{33}) only becomes exact in the limit $\ep\to0$. Indeed, in the presence of the regulator even the protected operator $D^2\cO_{\mathbf{20}'/6}$ has an ``anomalous dimension" $-2\ep$ corresponding to the regularized form of the tree graph.

Since we already know that $\cO_{\mathbf{20}'/6}$ (and consequently  $D^2\cO_{\mathbf{20}'/6}$) is a pure state normalized at unity, we can choose the mixing matrix in the following triangular form:
\begin{equation}\label{34}
  Z=
  \begin{pmatrix}
    Z_{\cK} & Z_{\cO} \\
    0 & 1
  \end{pmatrix} \ , \qquad
  Z^\dagger = Z^T =
  \begin{pmatrix}
    Z_{\cK} & 0 \\
    Z_{\cO} & 1
  \end{pmatrix} \ ,
\end{equation}
where we have taken into account the fact that the mixing coefficients are real.
We can derive differential equations for $Z_{\cK}$ and $Z_{\cO}$ by repeating the steps which lead to eq.\,(\ref{12}). Since we are now dealing with matrices, instead of taking the log of eq.\,(\ref{33}) we directly differentiate it. The derivative  $\pa/\pa g^2$ gives
\begin{equation}\label{35}
  \frac{\pa Z}{\pa g^2} G Z^\dagger + ZG\frac{\pa Z^\dagger}{\pa g^2} + x^{2\ep} Z \frac{\pa G}{\pa g^2x^{2\ep}} Z^\dagger  = O(1)
\end{equation}
(the details of the right-hand side do not matter for us). Further, differentiating with $x^2\pa/\pa x^2$ and dividing by $\ep g^2$ we obtain
\begin{equation}\label{36}
 x^{2\ep} Z \frac{\pa G}{\pa g^2x^{2\ep}} Z^\dagger = -\frac{\gamma}{g^2\ep}\begin{pmatrix}
     C_{\cK}(x^2)^{-\gamma(g^2)} & 0 \\
    0 & 0
  \end{pmatrix} + O(1)\,.
\end{equation}
We can now use this to rewrite (\ref{35}) in the form
\begin{equation}\label{37}
  \frac{\pa Z}{\pa g^2} Z^{-1} (Z G Z^\dagger) + (Z G Z^\dagger) (Z^\dagger)^{-1}\frac{\pa Z^\dagger}{\pa g^2} - \frac{\gamma}{g^2\ep} \begin{pmatrix}
     C_{\cK}x^{2\gamma} & 0 \\
    0 & 0
  \end{pmatrix}  = O(1)\,.
\end{equation}
Finally, substituting the initial equation (\ref{33}) into (\ref{37}) we find
\begin{equation}\label{38}
  \frac{\pa Z}{\pa g^2} Z^{-1} \begin{pmatrix}
    C_{\cK}x^{2\gamma} & 0 \\
    0 & 1
  \end{pmatrix} + \begin{pmatrix}
    C_{\cK}x^{2\gamma} & 0 \\
    0 & 1
  \end{pmatrix}  (Z^\dagger)^{-1}\frac{\pa Z^\dagger}{\pa g^2} + \frac{\gamma}{g^2\ep}\begin{pmatrix}
     C_{\cK}x^{2\gamma} & 0 \\
    0 & 0
  \end{pmatrix}  = O(1)\,.
\end{equation}

Solving this matrix differential equation is greatly simplified due to the triangular structure (\ref{34}). Inserting it into (\ref{38}) and dividing by the finite factor $C_{\cK}x^{2\gamma}$, we obtain two differential equations:
\begin{eqnarray}
  && Z'_{\cK} Z^{-1}_{\cK} - \frac{\gamma}{2g^2\ep}  = O(1)\,; \label{39}\\
  && Z'_{\cO} - (Z'_{\cK} Z^{-1}_{\cK})Z_{\cO}  = O(1)\,, \label{40}
\end{eqnarray}
where $Z'$ denotes the derivative with respect to $g^2$. Here we need to make the following comment. The form of the matrices appearing in (\ref{38}) is not exact. For example, the zeros in them are in fact $O(\ep)$ terms. Since the equation involves the inverse of the singular matrix $Z$, we should be careful not to make a mistake because of superposition of poles and $O(\ep)$ terms. A closer look shows that this does not happen. For instance, the $O(\ep)$ terms in the upper right corners of these matrices induce a shift of $Z_{\cO}$ in (\ref{40}),
\begin{equation}\label{41}
  Z'_{\cO} - (Z'_{\cK} Z^{-1}_{\cK})(Z_{\cO} + O(\ep))  = O(1)\,.
\end{equation}
Our aim here is to determine just the divergent and the finite parts of $Z_{\cO}$, so such a shift does not matter for us.

Equation (\ref{39}) is identical with (\ref{12}) and has the same solution (\ref{13}), provided we fix the finite normalization by the same requirement that $Z_{\cK}$ start with 1 and that it contain only poles (minimal subtraction); in this case the right-hand side of eq.\,(\ref{39}) vanishes. Replacing $Z'_{\cK} Z^{-1}_{\cK}$ in (\ref{40}) we obtain an equation for $Z_{\cO}$:
\begin{equation}\label{42}
  Z'_{\cO} - \frac{\gamma}{2g^2\ep} Z_{\cO} = O(1) \,.
\end{equation}
At first sight it seems that this equation is the same as (\ref{39}), so it should have the same solution. In fact, this is not true because of the presence of finite terms in $Z_{\cO}$. Indeed, we already know that $Z_{\cO}$ does not start with 1 like $Z_{\cK}$, but with $g^2$. This follows form the form of the one-loop ``Konishi anomaly" (\ref{07'}), after switching from $\cK_{\mathbf{10}/1}$ to $\cK_{\mathbf{10}/6}$. For $g=0$ both the $B$ and the $F$ terms in (\ref{07'}) are absent, and the Konishi multiplet becomes ``semishort", $\bD^2\cK=0$.  The numerical factor $N/32\pi^2$ in (\ref{07'}) is the first term in the expansion of $Z_{\cO}$. Thus, if the boundary condition for $Z_{\cK}$ was $Z_{\cK}(g=0)=1$, for $Z_{\cO}$ it is $Z_{\cO}(g=0)=0$. Further, when solving for $Z_{\cK}$ we declared that its finite part was 1, which can always be achieved by a suitable overall finite renormalization of $\cK_{\mathbf{10}/6}$. Once this has been done, we do not have any freedom left to fix the finite part of $Z_{\cO}$, so we must keep it arbitrary in the right-hand side of (\ref{42}).

So, let us write down
\begin{equation}
Z_{\cO}(g^2,\epsilon)= Z_{\cO s}(g^2,\epsilon)  + Z_{\cO f}(g^2) +
O(\epsilon)
 \label{ZO}
\end{equation}
where $Z_{\cO s}$ is the singular in $\epsilon$ part, while $Z_{\cO f}$
is the finite part. Then eq.\,(\ref{42}) becomes
\begin{equation}
Z_{\cO s}' - {\gamma(g^2) \over 2 g^2 \epsilon} Z_{\cO s} = {\gamma(g^2) \over 2 g^2 \epsilon} Z_{\cO f} + O(1)
 \label{new48}
\end{equation}
with the boundary condition $Z_{\cO s}(0,\epsilon) = 0 $. It is easy to check that the solution of (\ref{new48}) has the following form:
\begin{equation}\label{43}
  Z_{\cO s}(g^2,\epsilon) = - Z_{\cK}(g^2) \int_{0}^{g^2} Z_{\cO f}(\tau) (Z_{K}^{-1})^{\prime} (\tau)
 d\tau \,.
\end{equation}
In this equation $Z_{\cK}$ is the renormalization factor of the Konishi operator
(\ref{13}), and the finite part $Z_{\cO f}(g^2) = 0 + \omega_0 g^2 + \omega_1 g^4 + \omega_2 g^6 + \ldots$ should be determined by an explicit graph calculation. Let us also give the first few terms in the perturbative expansion of $Z_{\cO}$:
\begin{equation}\label{43''}
  Z_{\cO}(g^2,\epsilon) = \omega_0 g^2 + \left(\omega_1 + \frac{\omega_0\gamma_1}{4\ep}  \right)g^4 +  \left(\omega_2 + { \omega_1 \gamma_1 \over 6 \epsilon} + \frac{ \omega_0 \gamma_2}{ 6 \epsilon} + \frac{ \omega_0 \gamma_1^2}{24 \epsilon^2}    \right)g^6 +
O(g^8)\,.
\end{equation}

In conclusion we can say that we have resolved the mixing of the operators $B_6$ and $D^2\cO_{\mathbf{20}'/6}$. We have found two orthogonal pure states, $\cK_{\mathbf{10}/6}$ and $D^2\cO_{\mathbf{20}'/6}$. The entire information about the mixing coefficients is encoded in two quantities, the anomalous dimension of the Konishi multiplet $\gamma$ and the finite part of $Z_{\cO}$. The former has a well-defined meaning in a conformal theory and is guaranteed to be scheme independent. The latter, however, does not have any scheme-independent meaning at all. We might pull out the common renormalization factor $Z_{\cK}$ and then identify the so-called ``Konishi anomaly" with the remaining factor $-\int_{0}^{g^2} Z_{\cO f}(\tau) (Z_{K}^{-1})^{\prime} (\tau)$ in front of $D^2\cO_{\mathbf{20}'/6}$. At the lowest order $g^2$ it reproduces the ``one-loop Konishi anomaly", i.e., the numerical factor in (\ref{07'}). Beyond this order, this factor becomes singular. Attempting to identify the ``Konishi anomaly" with the finite part of $Z_{\cO}$ does not make much more sense, since it obviously is scheme dependent. Thus, we are forced to conclude that, in the context of the superconformal $\cN=4$ theory, the so-called ``Konishi anomaly" is nothing but a (divergent) mixing coefficient which may receive corrections at every loop order.

\subsection{The superdescendant $\cO_{\mathbf{10}/6}$}

While discussing the mixing problem above, we only dealt with the superdescendant $\cK_{\mathbf{10}/6}$ of the Konishi multiplet and kept that of the half-BPS operator $\cO_{\mathbf{20}'}$ implicit, in the form of the supercomponent $D^2\cO_{\mathbf{20}'/6}$ rather than the explicit mixture
\begin{equation}\label{44}
 D^2\cO_{\mathbf{20}'/6}   = \cZ_F F_6 + \cZ_B B_6 = \cO_{\mathbf{10}/6} \,.
\end{equation}
This is the generalization of the naive (classical) expression (\ref{015}), taking into account the possibility of quantum corrections. In this section we analyze the two-point functions of $\cO_{\mathbf{10}/6}$ with $D^2\cO_{\mathbf{20}'/6}$ and $\cK_{\mathbf{10}/6}$    up to order $g^4$ and find some restrictions on the renormalization factors $\cZ_F,\cZ_B$.

Let us start with the correlator
\begin{equation}\label{45}
  \langle \bD^2\bar \cO_{\mathbf{20}'/6} \cO_{\mathbf{10}/6}\rangle = \cZ_F \langle \bD^2\bar \cO_{\mathbf{20}'/6} F_6\rangle + \cZ_B \langle \bD^2\bar \cO_{\mathbf{20}'/6} B_6\rangle = \langle \bD^2\bar \cO_{\mathbf{20}'/6} D^2\cO_{\mathbf{20}'/6}\rangle = 1 + O(\ep)\,,
\end{equation}
which is a component/descendant of the protected correlator $\langle\bar\cO_{\mathbf{20}'/6} \cO_{\mathbf{20}'/6}\rangle = 1 + O(\ep)$.
The two-point functions $\langle \bD^2\bar \cO_{\mathbf{20}'/6} B_6\rangle$ (recall (\ref{31})) and $\langle \bD^2\bar \cO_{\mathbf{20}'/6} F_6\rangle$ have the following perturbative expansions:
\begin{eqnarray}
  \langle \bD^2\bar \cO_{\mathbf{20}'/6} B_6\rangle &=&  g^2(x^2)^{-3+3\ep}[b_0(\ep) + g^2(x^2)^{\ep} b_1(\ep)+ \ldots] \label{46'}\\
  \langle \bD^2\bar \cO_{\mathbf{20}'/6} F_6\rangle  &=&  (x^2)^{-3+2\ep}[1 + g^2(x^2)^{\ep} f_1(\ep)+ g^4(x^2)^{2\ep} f_2(\ep)+  \ldots]\,. \label{46}
\end{eqnarray}
Here the coefficients have a pole structure according to the general pattern (\ref{6}):
\begin{eqnarray}
  && b_0(\ep) = b_{00}+ b_{0,-1}\ep + O(\ep^2) \nonumber\\
  && b_1(\ep) = \frac{b_{11}}{\ep} + b_{10}+ O(\ep) \nonumber\\
  && f_1(\ep) = \frac{f_{11}}{\ep} + f_{10}+ O(\ep) \nonumber\\
  && f_2(\ep) = \frac{f_{22}}{\ep^2} + \frac{f_{21}}{\ep} + f_{20}+ O(\ep)\label{47}
\end{eqnarray}
which  follows from analyzing the corresponding graphs. In order to compensate these poles and  to assure the standard normalization at unity in (\ref{45}), we introduce the renormalization factors $\cZ_F,\cZ_B$ of the form
\begin{eqnarray}
  &&\cZ_F = 1 + g^2 \left(\frac{\vp_{11}}{\ep} + \vp_{10}  \right) + g^4 \left(\frac{\vp_{22}}{\ep^2} + \frac{\vp_{21}}{\ep} + \vp_{20}   \right) + O(g^6) \nonumber\\
  &&\cZ_B = 1 + g^2 \left(\frac{\beta_{11}}{\ep} + \beta_{10}  \right)+ O(g^4) \label{48}\,.
\end{eqnarray}

Now, let us put all this in the relation (\ref{45}) and expand  in $g^2$. The condition we find at level $g^2$ is
\begin{equation}\label{49}
  \frac{\vp_{11}}{\ep} + \vp_{10} + [f_1(\ep) + b_0(\ep)](x^2)^{\ep} = O(\ep)\,.
\end{equation}
The expansion in $\ep$ gives three types of terms, $\ep^{-1}$, $\ep^0\ln x^2$ and $\ep^0$, which are not present in the right-hand side of (\ref{49}) and so must vanish. This implies
\begin{equation}\label{50}
  \vp_{11} = f_{11} = 0\,, \qquad \vp_{10}+f_{10}+b_{00}=0\,.
\end{equation}
At level $g^4$, using (\ref{50}), we find
\begin{equation}\label{51}
  \frac{\vp_{22}}{\ep^2} + \frac{\vp_{21}}{\ep} + \vp_{20} + \left(\frac{\beta_{11}}{\ep} + \beta_{10}  \right) b_0(\ep)(x^2)^{\ep} + [f_2(\ep) + b_1(\ep)](x^2)^{2\ep} = O(\ep)\,.
\end{equation}
From the vanishing of the terms $\ep^{-2}$ and $\ep^{-1}\ln x^2$ we obtain
\begin{equation}\label{52}
  \vp_{22} = f_{22} = 0\,.
\end{equation}
Similarly, the terms $\ep^{-1}$ and $\ep^0\ln x^2$ yield
\begin{equation}\label{53}
  \vp_{21} = f_{21} + b_{11} = -\frac{1}{2}\beta_{11} b_{00}\,.
\end{equation}

The conclusion from the analysis of the protected two-point function (\ref{45}) is that the leading poles in $\cZ_F$ must be absent, whereas the subleading pole at level $g^4$ is related to the level $g^2$ pole in $\cZ_B$. The latter can be determined form the orthogonality condition
\begin{equation}\label{54}
  \langle \bar\cK_{\mathbf{10}/6} \cO_{\mathbf{10}/6}\rangle = O(\ep)\,.
\end{equation}
Substituting the definitions (\ref{020}) and (\ref{44}) into (\ref{54}) and using (\ref{45}), we can write down
\begin{equation}\label{55}
  Z_{\cK}\cZ_B \langle \bar B_6B_6 \rangle + Z_{\cK}\cZ_F \langle \bar B_6F_6 \rangle + Z_{\cO}(1+O(\ep)) = O(\ep)\,.
\end{equation}

We want to study the consequences of this condition up to level $g^4$. The relevant two-point functions are
\begin{eqnarray}
  \langle \bar B_6B_6\rangle &=&  g^2 (x^2)^{-3+3\ep}[1 + g^2(x^2)^{\ep} a_1(\ep) + \ldots] \label{56'}\\
  \langle \bar B_6F_6 \rangle &=&  g^2 (x^2)^{-3+3\ep}[d_0(\ep) + g^2(x^2)^{\ep} d_1(\ep)+ \ldots]\,, \label{56}
\end{eqnarray}
where $a_1(\ep) = a_{11}\ep^{-1} + a_{10} + O(\ep)$, $d_0(\ep) = d_{00} + d_{0,-1}\ep + O(\ep^2)$ and $d_1(\ep) = d_{11}\ep^{-1} + d_{10} + O(\ep)$. Further, the renormalization factors $Z_{\cK}, Z_{\cO}$ were discussed in Section \ref{s3}, where we found the following (recall (\ref{13}) and (\ref{43''})):
\begin{equation}\label{57}
  Z_{\cK} = 1 + g^2 \frac{\gamma_1}{2\ep} + O(g^4)\,, \qquad Z_{\cO} = g^2 \omega_{0} + g^4 \left(\frac{\omega_{0}\gamma_1}{4\ep} + \omega_1 \right) + O(g^6)\,.
\end{equation}

Let us now substitute all this into the orthogonality condition (\ref{55}) and expand up to level $g^4$. We note that the $O(\ep)$ term multiplying $Z_{\cO}$ is of order $g^2$, and so is the non-singular part of $Z_{\cO}$, so this term has no effect at the levels we are interested in. Thus, at level $g^2$ we find
\begin{equation}\label{59}
  \omega_{0} + [1+d_0(\ep)](x^2)^{\ep} = O(\ep) \ \Rightarrow \ \omega_{0}+d_{00}+1 = 0\,.
\end{equation}
The condition at level $g^4$ is
\begin{equation}\label{60}
  \frac{\omega_{0}\gamma_1}{4\ep} + \omega_1 + \left[\frac{\gamma_1+2\beta_{11}}{2\ep} + \beta_{10} +\left(\frac{\gamma_1}{2\ep} +\vp_{10}  \right)d_0(\ep)  \right](x^2)^{\ep} + \left[a_1(\ep) + d_1(\ep)\right](x^2)^{2\ep} = O(\ep)\,.
\end{equation}
Expanding in $\ep$, we find two conditions following from the vanishing of the terms $\ep^{-1}$ and $\ep^{0}\ln x^2$. The first of them is $a_{11}+d_{11}=\frac{\omega_{0}\gamma_1}{4}$, while the other determines the coefficient
\begin{equation}\label{61}
  \beta_{11} = 0\,.
\end{equation}
This result, together with (\ref{53}) and (\ref{52}), imply the vanishing of all the singular terms in both renormalization factors $\cZ_F,\cZ_B$.

Thus, the conclusion from our investigation is that the renormalization factors in (\ref{44}) can only contain finite terms:
\begin{equation}\label{63}
  \cZ_F = 1 + g^2 \vp_{10} + g^4 \vp_{20} + O(g^6)\,, \qquad \cZ_B = 1 + g^2\beta_{10}+ O(g^4)\,.
\end{equation}
These (possible) finite corrections to the naive descendant $\cO_{\mathbf{10}/6}$ (\ref{015}) obtained through use of the classical field equations can be determined from the $\ep^0$ parts of the above relations. One way to find $\beta_{10}$ is to examine the $\ep^0$ terms in (\ref{60}), but it is more efficient to go back to the starting point, the expected operator relation (\ref{44}) and apply it on the two-point functions with the operator $B_6$:
\begin{equation}\label{621}
  \langle \bD^2\bar \cO_{\mathbf{20}'/6} B_6\rangle = \cZ_F \langle \bar F_6 B_6\rangle + \cZ_B \langle \bar B_6 B_6\rangle\,.
\end{equation}
Substituting in it the expansions (\ref{46'}), (\ref{56'}) and (\ref{56}) and working out the $\ep^0$ terms, we find the relation
\begin{equation}\label{622}
  b_{10} = d_{10} + a_{10} + \vp_{10} d_{00} + \beta_{10}\,,
\end{equation}
which determines $\beta_{10}$. The other finite correction, $\vp_{20}$, can be obtained from the $\ep^0$ terms in (\ref{51}):
\begin{equation}\label{53'}
  \vp_{20} + \vp_{10}f_{10} + f_{20} + \beta_{10}b_{00}+b_{10} =0\,.
\end{equation}

\section{Explicit calculations up to order $g^4$}\label{se3}

\subsection{The operator identity between $D^2 {\cal O}, F, B$} \label{opid}

For simplicity the $SU(4)$ and $SU(3)$ labels are omitted throughout this section; we discuss
the mixing problem in the $\mathbf{6}$ of $SU(3)$. Recall that the
naive application of the equation of motion of the chiral superfield
implies the operator identity:
\begin{equation}
D^2 {\cal O} \, + \, \frac{1}{2} \, (F \, - \, 4 \, B) \, = \, 0\,.
\label{o10eq}
\end{equation}
We argue that this equation holds
for the \emph{bare} operators, i.e. that there is no need to introduce
renormalization factors. The analysis of the last section excluded infinite
renormalization factors, here we show that in the SSDR scheme even
finite renormalization effects are absent. Further, in this scheme the
right-hand side of equation (\ref{o10eq}) is exactly zero (i.e., not only in the limit $\ep\to 0$).

In the Appendix we comment on the calculation of the correlation functions
employed here. We remark that the considerations in this subsection are based
on the use of ${\cal N} = 1$ superpropagators and partial integration, while
the actual regularization is of relevance only with respect to  one detail,
namely eq.\,(\ref{eqreg}) which is certainly true in dimensional regularization.
%Unfortunately, we cannot prove it in general.

In order to establish the operator relation (\ref{o10eq}), we have to verify the vanishing of the two-point functions
of the linear combination on the left-hand side of eq.\,(\ref{o10eq}) with each of the operators
$F, \, D^2 {\cal O}$ and $B$. To lowest order we find
\begin{eqnarray}
&& \langle F \, \bar F \rangle_{g^0} \, = \,
-2 \, \langle D^2 {\cal O} \, \bar F \rangle_{g^0}\, = \,
4 \, \langle D^2 {\cal O} \, \bar D^2 \bar{\cal O} \rangle_{g^0}
\nonumber \\ && = - 16 (N^2-1) (\partial_1^\mu \Pi_{12}) (\partial_{1\mu}
\Pi_{12}) \, ,
\end{eqnarray}
with the matter propagator
\begin{equation}
\Pi_{12} \, = \, \frac{c_0}{x_{12}^2} \, , \qquad c_0 = -\frac{1}{4\pi^2}\,.
\end{equation}
Hence equation (\ref{o10eq}) is satisfied.

At order $g^2$ we first consider the two-point function with $\bar B$. We have
\begin{equation}
\langle B \, \bar B \rangle_{g^2} = 2g^2 N (N^2-1) \,
\Pi_{12}^3 \,
\end{equation}
(recall that the definition (\ref{010}) of $B$ includes a factor of $g$), while
\begin{equation}
\langle D^2 {\cal O} \, \bar B \rangle_{g^2} \, = \, 4 g^2 N
(N^2-1) \, \Pi_{12}^3 \, \, . \label{OBg1}
\end{equation}
The latter correlator involves only one diagram
which is rational and finite. Since $F$ and $B$ do not couple at this order,
the coefficient of $B$ in (\ref{o10eq}) is confirmed.

Next, we test eq.\,(\ref{o10eq}) with respect to  $F$ at order $g^2$:
\begin{equation}
\langle F \, \bar F \rangle_{g^2} \, = \, -
2 \, \langle D^2 {\cal O} \, \bar F \rangle_{g^2} \, = \, - 32 g^2 N
(N^2-1) \, \bigl( \, \Pi_{12}^3 \, - \frac{1}{2} \square_1 \square_1 \,
f(1,2;1,2) \bigr)\, .
\end{equation}
Here we have introduced the double-box integral
\begin{equation}
f(1,2;1,2) \, = \, c_0^5 \, \int \frac{d^4x_5 \, d^4x_6}{x_{15}^2 x_{16}^2
x_{56}^2 x_{25}^2 x_{26}^2}\, .
\end{equation}
It is convergent and of dimension two, hence it is proportional to $1/x^2_{12}$. A box operator makes it into a
contact term. The same contact part occurs in $4 \, \langle D^2
{\cal O} \, \bar D^2 \bar {\cal O} \rangle_{g^2}$, which, however,
has no finite part. Equation (\ref{o10eq}) remains true
also in a two-point function with respect to  $D^2 {\cal O}$ because there is a
second finite contribution from $\langle B \, \bar D^2 \bar{\cal O}
\rangle_{g^2}$ (see (\ref{OBg1})).

Let us finally consider a test of (\ref{o10eq}) with respect to  $B$ at order $g^4$.
We find
\begin{eqnarray}
\langle D^2 {\cal O} \, \bar B \rangle_{g^4} & = & 12 g^4 N^2 (N^2-1) \,
\bigl(J \, - \, 2 \, \Pi_{12} \, h + \Pi_{12} \, \square_1 \, f \bigr) \, ,
\nonumber \\ \langle F \, \bar B \rangle_{g^4} & = & 24 g^4 N^2 (N^2-1) \,
( - J ) \, , \\ \langle B \, \bar B \rangle_{g^4} & = & \phantom{2}6
g^4 N^2 (N^2-1) \, \bigl( - 2 \, \Pi_{12} \, h + \Pi_{12} \, \square_1 \,
f \bigr) \, ,
\nonumber
\end{eqnarray}
where $f \, = \, f(1,2;1,2)$ and
\begin{eqnarray}
h & = & c_0^4 \, \int \frac{d^4x_5}{x_{15}^4 x_{25}^4} \, , \\
J & = & c_0^7 \, \int \frac{d^4x_{3,5,6} \, (\partial_{15} \, \partial_{35}
\, \partial_{36} \, \partial_{16})}{x_{15}^2 x_{35}^2
x_{36}^2 x_{16}^2 x_{25}^2 x_{23}^2 x_{26}^2} \, .
\end{eqnarray}

Once again, up to $O(g^4)$ equation (\ref{o10eq}) is seen to be identically
satisfied. In showing this we did not rely on the explicit evaluation of
any divergent $x$-space integral.

We conjecture that the operator relation is true to all orders in perturbation
theory without receiving any modification of its coefficients. In conclusion,
\begin{equation}
- 2 \, D^2 \, {\cal O} \, = \, F \, - \, 4  \, B \, \equiv \, [F]_R
\end{equation}
i.e. we view the combination $F - 4 B$ as the correctly renormalised
operator $F$. In the following we choose to drop the bare operator $F$ from our mixing
problem in favour of $D^2 {\cal O}$.

\subsection{$\cK_{\mathbf{10}}$ by orthogonalization}

Let us introduce the shorthand
\begin{eqnarray}
\langle D^2 {\cal O} \, \bar B \rangle & = & g^2 \, OB_1 \, + \,
g^4 \, OB_2 \, + \, \ldots \, , \\ \langle D^2 {\cal O} \, \bar D^2
\bar {\cal O} \rangle & = & g^0 \, OO_0 \, + \,  g^2 \, OO_1 \, + \,
\ldots \nonumber \, .
\end{eqnarray}
We are looking for a combination
\begin{equation}
\cK_{\mathbf{10}} \, = \, Z_{\cal K} \, B \, + Z_{\cal O} \, D^2 {\cal O}
\end{equation}
orthogonal to $D^2 {\cal O}$ up to $O(g^4)$. The relevant part of
the $Z$-factors is
\begin{eqnarray}
Z_{\cal K} & = & 1 + g^2 \frac{m_{11}}{\epsilon} \label{zfactors} \\
Z_{\cal O} & = & g^2 n_{10} + g^4 \left( \frac{n_{21}}{\epsilon} +
n_{20} \right)\,, \nonumber
\end{eqnarray}
i.e. we take $Z_{\cal K}$ to have minimal subtraction form. We impose
\begin{equation}
\langle \cK_{\mathbf{10}} \, \bar D^2 \bar{\cal O} \rangle \, = \, O(\epsilon)
\, .
\end{equation}
At order $g^2$ this yields one equation from the elimination of the finite
part, whereas at order $g^4$ we generate three equations relating to the
elimination of the simple pole, the simple logarithm and the finite part.
The system is non-singular so that the four constants in the
$Z$-factors are uniquely determined. \\

The integrals $J,h,f$ can be calculated in $p$-space by the {\it Mincer} package
\cite{Verma}. We Fourier-transform back to $x$-space to find
\begin{eqnarray}
OO_0 & = & - 16 (1 - 2 \epsilon + \epsilon^2) (x^2)^{(2 \epsilon)} \, , \\
OO_1 & = & - 8 M (0 + 12 \zeta(3) \epsilon + \ldots) (x^2)^{(3 \epsilon)} \, ,
\nonumber \\ OB_1 & = & - 4 M (x^2)^{(3 \epsilon)}
\, , \nonumber \\ OB_2 & = & + 3 M^2 (1/\epsilon + 3 + (8 + 12 \zeta(3))
\epsilon + \ldots) (x^2)^{(4 \epsilon)} \, ,
\nonumber
\end{eqnarray}
where an overall factor
\begin{equation}
c_1 \, = \, \frac{(N^2-1)}{(4 \pi^2)^2 x^6_{12}}
\end{equation}
has been omitted and $M = N/(4 \pi^2)$. For the details on our convention
we refer to \cite{EJS}. The expressions above are valid in the improved
$\overline{MS}$ scheme, in which fractional powers of $\pi$, the Euler constant
$\gamma$ and $\zeta(2)$ are absorbed into the mass scale.

The result of the orthogonalization is:
\begin{eqnarray}
m_{11} = \frac{3 M}{2} \, , \qquad n_{10} = -
\frac{M}{4} \, , \qquad n_{21} = - \frac{3 M^2}{16} \, ,
\qquad n_{20} = + \frac{3 M^2}{16} \, .
\end{eqnarray}
The explicit calculation is needed, of course, to furnish the principle pieces
of information, i.e. the anomalous dimensions and the finite mixing
coefficients $\omega_{i}$. Nevertheless, in the context of this
paper we rather view it as a confirmation of the abstract analysis presented
in the earlier sections, and thus as a demonstration of superconformal
invariance. The consistency conditions on the coefficients in the various
correlators are in fact fulfilled in calculations in SSDR, notably
\begin{equation}
m_{11} = \frac{\gamma_1}{2}
\end{equation}
and
\begin{equation}
n_{21} \, = \, \frac{1}{2} \, m_{11} \, n_{10} \, ,
\end{equation}
so that $Z_{\cal O}$ is definitely singular. This had already been derived
in the preceding sections.

\section*{Acknowledgments} We profited a lot from enlightening discussions with D. Anselmi, G. Arutyunov, M. Grisaru, J. Iliopoulos, H. Osborn, G.C. Rossi, K. Stelle, R. Stora, A. Vainshtein. ES is grateful to the theory group of the Dipartimento di Fisica, Universit\`a di Roma ``Tor Vergata" and to the Albert-Einstein Institut, Potsdam  for the warm hospitality extended to him. The work of E.S. was supported in part by the INTAS contract 00-00254 and by the MIUR-COFIN contract 2003-023852. The work of Ya.S.S. was supported in part by the INFN, by the MIUR-COFIN contract 2003-023852, by the EU contracts MRTN-CT-2004-503369 and MRTN-CT-2004-512194,
by the INTAS contract 03-51-6346 and by the NATO grant PST.CLG.978785.

\section*{Appendix: Some comments on the graphs}

A complete set of conventions with respect to spinor algebra, Fourier
transform and regularization by supersymmetric dimensional reduction (SSDR)
is given in \cite{EJS}. For convenience we quote a few formulae. The
classical action of ${\cal N}=4$ SYM in terms of ${\cal N} = 1$ fields is
\begin{eqnarray}
  S_{\cN=4}&=& \int d^4x \, d^2\theta \, d^2\bar\theta\; \T \left(e^{gV} \bar
\Phi_I e^{-gV} \Phi^I  \right) \label{67}  \\
  &+&  \frac{1}{4}\int d^4x_L \, d^2\theta\; \T(W^\alpha W_\alpha) + \left[
\frac{g}{3!}\int d^4x_L \, d^2\theta\; \ep_{IJK} \T (\P^I[\P^J,\P^K]) + c.c.
\right]\,, \nonumber
\end{eqnarray}
where all superfields are in the adjoint of $SU(N_c)$ and the generators are
normalised such that $\T(T^a T^b) \, = \, \delta^{ab}$. The matter propagator
is
\begin{equation}
\langle \Phi^I(1) \bar \Phi_J(2) \rangle \, = \,
 - \frac{\delta^I_J }{4 \pi^2 \hat x_{12}^2} \label{71} \, .
\end{equation}
The hatting indicates supersymmetrization by the exponential shift
\begin{equation}\label{721}
\frac{1}{\hat x_{12}^2} \, = \, \exp\left\{i \,[\theta_1 \sigma_\mu \bar
\theta_1  +  \theta_2 \sigma_\mu \bar \theta_2  - 2  \theta_1
\sigma_\mu  \bar \theta_2] \, \partial^\mu_1\right\} \frac{1}{x_{12}^2} \,
\equiv \, e^{\Delta_{12}} \frac{1}{x_{12}^2} \, .
\end{equation}
The Yang-Mills propagator in Feynman gauge is given by
\begin{equation}\label{73}
\langle V(1) V(2)\rangle = + \frac{\theta_{12}^2 \bar \theta_{12}^2}{4 \pi^2
x_{12}^2} \, .
\end{equation}
The SSDR prescription means to send
\begin{equation}
\frac{1}{x^2_{12}} \, \rightarrow \, \frac{1}{(x^2_{12})^{(1-\epsilon)}}
\end{equation}
while we absorb a corresponding change of the normalization of the propagators
into the mass scale (in the terminology of \cite{EJS} we use $\tilde \mu$
in $x$-space instead of $\mu$ itself). At the same time, the Grassmann variables are treated as two-component spinors, as in four dimensions.

The most efficient way of reducing supergraphs to ordinary integrals is to
expand in the $\theta$'s the exponential shifts from the matter propagators
so as to saturate the Grassmann integrations. After this we are left with
some differential operator acting under the $x$-integrals. For some simple
illustrations of the method see once again \cite{EJS}. \\

In the following we calculate with the operators in the \textbf{6} of $SU(3)$,
as defined in Section \ref{sec2}. Most conveniently one restricts to the 11
projection
\begin{equation}
B \, = \, g \, {\rm Tr} \left(\Phi^1 [\bar \Phi_2, \bar \Phi_3] \right) \, ,
\qquad F \, = \, {\rm Tr} \left(\nabla^\alpha \P^1 \nabla_\alpha \P^1 \right)
\, .
\end{equation}
Let us start with $\langle D^2 {\cal O} \, \bar B \rangle_{g^4}$, defined
by the graphs in Figure 1. The combinatorics produces
\begin{equation}
8 g^4 N^2 (N^2-1) \, \bigl( - \, G_8 + \, G_7 + \, G_6 + \frac{1}{2} \,
G_5 + \frac{1}{2} \, G_4 - \, G_3 - \, G_2 + \, G_1  \bigr) \, .
\end{equation}
Here the extra minus sign from the YM propagator has been included.

\vskip 0.1 cm

%---------- FIGURE TOP ------------
\begin{minipage}{\textwidth}
\begin{center}
\includegraphics[width=0.70\textwidth]{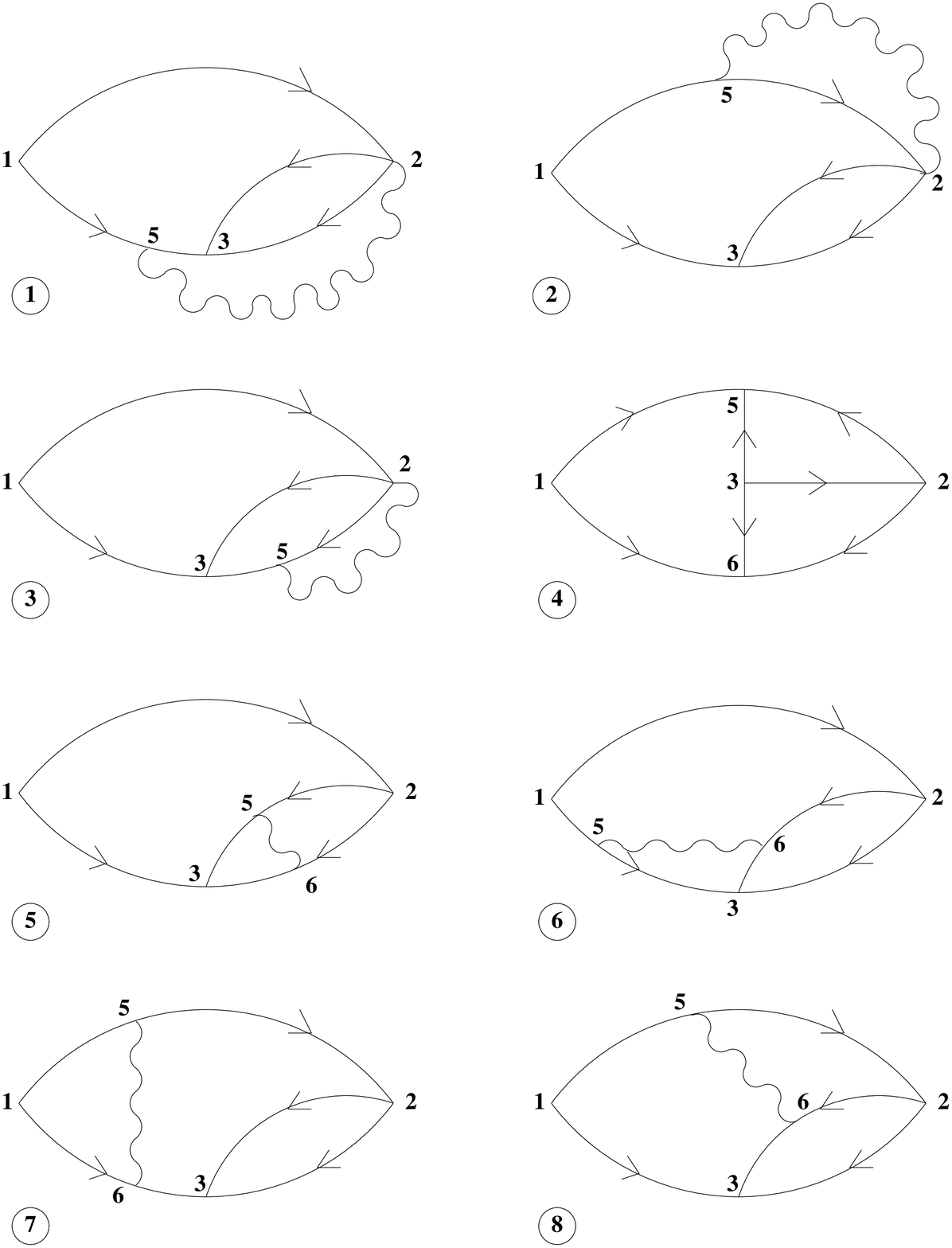}
\end{center}
\end{minipage}
\begin{center}
Figure 1. Graphs $G_{1\ldots8}$ for the correlator $\langle {\cal O}_{\mathbf{20}/6} \, \bar
B \rangle_{g^4}$.
\end{center}
%---------- FIGURE END ------------

\vskip 0.2 cm

By means of partial integration in $x$-space and
shrinking of some lines onto which box operators act, one may show
\begin{equation}
G_8 \, = \, G_7 + G_6 - \, G_5 - \, G_4 - \, G_3 - \, G_2 \, ,
\end{equation}
whereas graph $G_1$ vanishes. We conclude:
\begin{equation}
\langle D^2 {\cal O} \, \bar B \rangle_{g^4} \, = \,
12 g^4 N^2 (N^2-1) \, \bigl( G_4 + G_5 \bigr)  \label{OBg3}
\end{equation}
The supergraphs $G_4$ and $G_5$ yield the underlying $x$-space integrals
\begin{eqnarray}
G_4 & = & J \, - 2 \, \Pi_{12} \, h \, , \\
G_5 & = & \Pi_{12} \, \square_1 f(1,2;1,2)   \nonumber
\end{eqnarray}
(the functions $J,h,f$ were defined in Section \ref{opid}).

The calculation of $\langle F \, \bar B \rangle_{g^4}$ is very similar. The part of this correlator arising from the two-fermion term in $F$ can be
read off from the calculation of $\langle O \, \bar B \rangle_{g^4}$: we
simply drop those parts of the graphs in which both derivatives at point
1 act on the same line. In the YM sector only graphs $G_8$ and $G_7$ survive;
the cancellation of diagram $G_7$ against a part of diagram $G_8$ is not
affected. We find
\begin{equation}
- G_8 \, + \, G_7 \, = 8 g^4 N^2 (N^2-1) \bigl( - 2 \, J + 2 \, \Pi_{12} \,
h \bigr) \, .
\label{cancelFC}
\end{equation}
Next, there are six graphs in which a connection line emanates
from $F$. Five of these vanish by $\theta$-counting, the remaining diagram
is displayed in Figure 2. It contributes
\begin{equation}
G_9 \, = \, 8 g^4 N^2 (N^2-1) \, (- 2 \, \Pi_{12} \, h) \, ,
\end{equation}
which exactly compensates the $h$ term in (\ref{cancelFC}). Finally, the
matter sector graph $G_4$ comes without the $h$-pieces that it had before. On
adding up we obtain:
\begin{equation}
\langle F \, \bar B \rangle_{g^4} \, = \, 24 g^4 N^2 (N^2-1) \,
(-J) \, .
\end{equation}

\vskip 0.3 cm

%---------- FIGURE TOP ------------
\begin{minipage}{\textwidth}
\begin{center}
\includegraphics[width=0.40\textwidth]{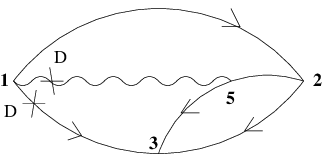}
\end{center}
\end{minipage}
\begin{center}
Figure 2. The graph $G_9$ in the correlator $\langle F \, \bar B \rangle_{g^4}$.
\end{center}
%---------- FIGURE END ------------

\vskip 0.2 cm

The graphs contributing to $\langle D^2 {\cal O} \bar F
\rangle_{g^2}$ are given in Figure 3. Graph $G_{II}$ is zero, while the other
two graphs can rather straightforwardly be summed into
\begin{equation}
\langle D^2 {\cal O} \, \bar F \rangle_{g^2} \, = \, 16 g^2 N
(N^2-1) \, \bigl( \, \Pi_{12}^3 \, - \, \frac{1}{2}
\square_1 \square_2 \, f \bigr) \, .
\end{equation}
To show this it is enough to pull the two box operators in the last equation
through the integrations in $f$ and to distribute the derivatives on the
various propagators, although we rely on the fact that
\begin{equation}
\square_{15} \square_{25} \, f(1,2;1,2) \, = \, 0 \label{eqreg}
\end{equation}
in dimensional regularization. The last equation can be checked
in $p$-space from topology T1 in Mincer with numerator $p_1^2 \,
p_2^2$. Our operator equation (\ref{o10eq}) is actually valid
(up to $g^2$ in correlators with respect to $D^2 {\cal O}, \, F$ or $g^4$ with respect to $B$)
in all regularization schemes that are compatible with partial integration
and in which eq.\,(\ref{eqreg}) holds. \\

The evaluation of $\langle F \, \bar F \rangle_{g^2}$ is not essentially
different, while the $O(g^2)$ two-point function of $D^2 {\cal O}$ with
itself is most conveniently done by considering $\langle O \, \bar O
\rangle_{g^2}$ and applying the derivatives afterwards.

\vskip 0.3 cm

%---------- FIGURE TOP ------------
\begin{minipage}{\textwidth}
\begin{center}
\includegraphics[width=0.70\textwidth]{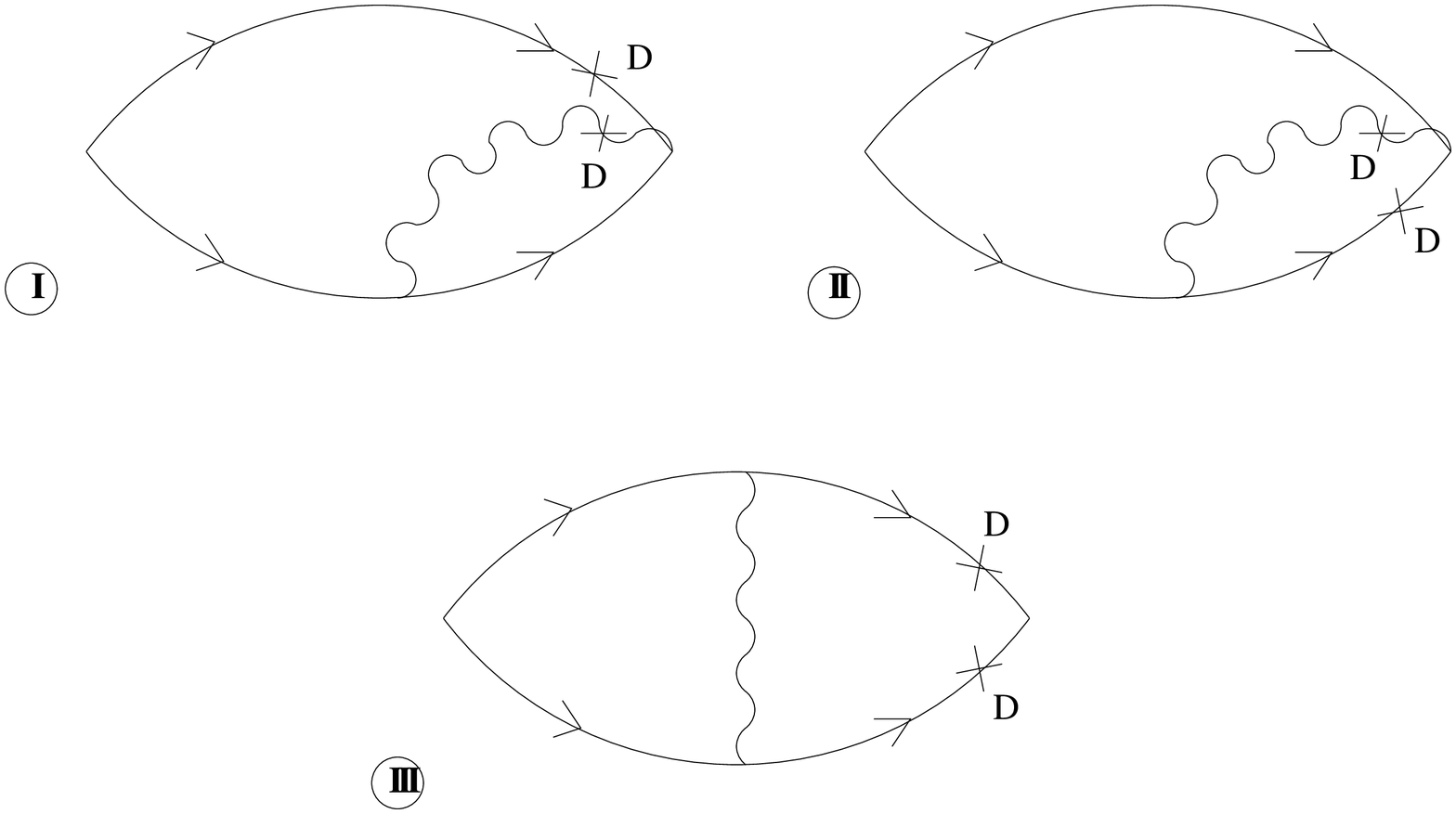}
\end{center}
\end{minipage}
\begin{center}
Figure 3. The correlator $\langle {\cal O}_{\mathbf{20}/6} \, \bar
F \rangle_{g^2}$.
\end{center}
%---------- FIGURE END ------------


\begin{thebibliography}{99}

\bibitem{PiguetKonishi}
T.~E.~Clark, O.~Piguet and K.~Sibold,
``Supercurrents, Renormalization And Anomalies,''
Nucl.\ Phys.\ B {\bf 143} (1978) 445;\\
K.~Konishi,
``Anomalous Supersymmetry Transformation Of Some Composite Operators In Sqcd,''
Phys.\ Lett.\ B {\bf 135} (1984) 439.

\bibitem{Grisaru} M.~T.~Grisaru, B.~Milewski and D.~Zanon,
``The Supercurrent And The Adler-Bardeen Theorem,''
Nucl.\ Phys.\ B {\bf 266} (1986) 589.


\bibitem{SV1} M.~A.~Shifman and A.~I.~Vainshtein,
``Solution Of The Anomaly Puzzle In Susy Gauge Theories And The Wilson
Operator Expansion,''
Nucl.\ Phys.\ B {\bf 277} (1986) 456
[Sov.\ Phys.\ JETP {\bf 64} (1986\ ZETFA,91,723-744.1986) 428].


\bibitem{EF} J.~Erlich and D.~Z.~Freedman,
``Conformal symmetry and the chiral anomaly,''
Phys.\ Rev.\ D {\bf 55} (1997) 6522
[arXiv:hep-th/9611133].

\bibitem{AGJ} D.~Anselmi, M.~T.~Grisaru and A.~Johansen,
``A Critical Behaviour of Anomalous Currents, Electric-Magnetic Universality
and CFT$_4$,''
Nucl.\ Phys.\ B {\bf 491} (1997) 221
[arXiv:hep-th/9601023].

\bibitem{SV2} M.~A.~Shifman and A.~I.~Vainshtein,
``Instantons versus supersymmetry: Fifteen years later,''
arXiv:hep-th/9902018.


\bibitem{CDSW} F.~Cachazo, M.~R.~Douglas, N.~Seiberg and E.~Witten,
``Chiral rings and anomalies in supersymmetric gauge theory,''
JHEP {\bf 0212} (2002) 071
[arXiv:hep-th/0211170].


\bibitem{Kon1}
D.~Anselmi, D.~Z.~Freedman, M.~T.~Grisaru and A.~A.~Johansen,
``Universality of the operator product expansions of SCFT(4),''
Phys.\ Lett.\ B {\bf 394} (1997) 329
[arXiv:hep-th/9608125].


\bibitem{Roma1} M.~Bianchi, S.~Kovacs, G.~Rossi and Y.~S.~Stanev,
``Anomalous dimensions in N = 4 SYM theory at order g**4,''
Nucl.\ Phys.\ B {\bf 584} (2000) 216
[arXiv:hep-th/0003203].

\bibitem{ESS} B.~Eden, C.~Schubert and E.~Sokatchev,
``Three-loop four-point correlator in N = 4 SYM,''
Phys.\ Lett.\ B {\bf 482} (2000) 309
[arXiv:hep-th/0003096]; ``Four-point functions of chiral primary operators in N = 4 SYM,''
arXiv:hep-th/0010005.

\bibitem{Arutyunov:2000im}
G.~Arutyunov, S.~Frolov and A.~Petkou,
``Perturbative and instanton corrections to the OPE of CPOs in N = 4
SYM(4),''
Nucl.\ Phys.\ B {\bf 602} (2001) 238
[Erratum-ibid.\ B {\bf 609} (2001) 540], hep-th/0010137.

\bibitem{Beisert:2003tq}
N.~Beisert, C.~Kristjansen and M.~Staudacher,
``The dilatation operator of N = 4 super Yang-Mills theory,''
Nucl.\ Phys.\ B {\bf 664} (2003) 131
[arXiv:hep-th/0303060].


\bibitem{Lipatov}
A.~V.~Kotikov, L.~N.~Lipatov, A.~I.~Onishchenko and V.~N.~Velizhanin,
``Three-loop universal anomalous dimension of the Wilson operators in N = 4
SUSY Yang-Mills model,''
Phys.\ Lett.\ B {\bf 595} (2004) 521
[arXiv:hep-th/0404092].

\bibitem{EJS}
B.~Eden, C.~Jarczak and E.~Sokatchev,
``A three-loop test of the dilatation operator in N = 4 SYM,''
arXiv:hep-th/0409009.

\bibitem{Berenstein:2002jq}
D.~Berenstein, J.~M.~Maldacena and H.~Nastase,
``Strings in flat space and pp waves from N = 4 super Yang Mills,''
JHEP {\bf 0204} (2002) 013
[arXiv:hep-th/0202021].

\bibitem{B}
B.~Eden,
``On two fermion BMN operators,''
Nucl.\ Phys.\ B {\bf 681} (2004) 195
[arXiv:hep-th/0307081].

\bibitem{Collins}
J.~C.~Collins,
``Renormalization. An Introduction To Renormalization, The Renormalization
Group, And The Operator Product Expansion,''  Cambridge, Uk: Univ. Pr. ( 1984) 380p.

\bibitem{Kelly}
P.~Breitenlohner, D.~Maison and K.~S.~Stelle,
``Anomalous Dimensions And The Adler-Bardeen Theorem In Supersymmetric
Yang-Mills Theories,''
Phys.\ Lett.\ B {\bf 134} (1984) 63.

\bibitem{Ensign:1987wy}
P.~Ensign and K.~T.~Mahanthappa,
``The Supercurrent And The Adler-Bardeen Theorem In Coupled Supersymmetric
Yang-Mills Theories,'' Phys.\ Rev.\ D {\bf 36} (1987) 3148.

\bibitem{Larin}
S.~A.~Larin,
``The Renormalization of the axial anomaly in dimensional regularization,''
Phys.\ Lett.\ B {\bf 303} (1993) 113
[arXiv:hep-ph/9302240].

\bibitem{I}
K.~A.~Intriligator and W.~Skiba,
``Bonus symmetry and the operator product expansion of N = 4
super-Yang-Mills,''
Nucl.\ Phys.\ B {\bf 559} (1999) 165
[arXiv:hep-th/9905020].

\bibitem{Roma2}
M.~Bianchi, S.~Kovacs, G.~Rossi and Y.~S.~Stanev,
``Properties of the Konishi multiplet in N = 4 SYM theory,''
JHEP {\bf 0105} (2001) 042
[arXiv:hep-th/0104016].


\bibitem{HH1} P.~Heslop and P.~S.~Howe,
``On harmonic superspaces and superconformal fields in four dimensions,''
Class.\ Quant.\ Grav.\  {\bf 17} (2000) 3743
[arXiv:hep-th/0005135].

\bibitem{FS} S.~Ferrara and E.~Sokatchev,
``Superconformal interpretation of BPS states in AdS geometries,''
Int.\ J.\ Theor.\ Phys.\  {\bf 40} (2001) 935
[arXiv:hep-th/0005151].

\bibitem{DF} E.~D'Hoker and D.~Z.~Freedman,
``Supersymmetric gauge theories and the AdS/CFT correspondence,''
arXiv:hep-th/0201253.


\bibitem{Penati}S.~Penati, A.~Santambrogio and D.~Zanon,
``Two-point functions of chiral operators in N = 4 SYM at order g**4,''
JHEP {\bf 9912} (1999) 006
[arXiv:hep-th/9910197].\\
S.~Penati, A.~Santambrogio and D.~Zanon,
``More on correlators and contact terms in N = 4 SYM at order g**4,''
Nucl.\ Phys.\ B {\bf 593} (2001) 651
[arXiv:hep-th/0005223].

\bibitem{DimRed} W.~Siegel,
``Supersymmetric Dimensional Regularization Via Dimensional Reduction,''
Phys.\ Lett.\ B {\bf 84} (1979) 193.\\
M.~T.~Grisaru, W.~Siegel and M.~Rocek,
``Improved Methods For Supergraphs,''
Nucl.\ Phys.\ B {\bf 159} (1979) 429.



\bibitem{Witten} E. Witten, ``Anti-de Sitter space and holography'',
Adv.Theor.Math.Phys.  {\bf 2} (1998) 253, hep-th/9802150.

\bibitem{HH2} P.~J.~Heslop and P.~S.~Howe,
``A note on composite operators in N = 4 SYM,''
Phys.\ Lett.\ B {\bf 516} (2001) 367
[arXiv:hep-th/0106238].



\bibitem{HSS}
A.~Galperin, E.~Ivanov, S.~Kalitsyn, V.~Ogievetsky and E.~Sokatchev,
``Unconstrained N=2 Matter, Yang-Mills And Supergravity Theories In Harmonic
Superspace,''
Class.\ Quant.\ Grav.\  {\bf 1} (1984) 469.\\
A.~S.~Galperin, E.~A.~Ivanov, V.~I.~Ogievetsky and E.~S.~Sokatchev,
``Harmonic superspace,''  Cambridge, UK: Univ. Pr. (2001) 306 p.

\bibitem{Verma} S.~A.~Larin, F.~V.~Tkachov and J.~A.~M.~Vermaseren,
``The FORM version of MINCER,''
NIKHEF-H-91-18.



\bibitem{AS}
G.~Arutyunov and E.~Sokatchev,
``A note on the perturbative properties of BPS operators,''
Class.\ Quant.\ Grav.\  {\bf 20} (2003) L123
[arXiv:hep-th/0209103].

\end{thebibliography}
\end{document}